\documentclass[twocolumn]{aastex631}

\usepackage{amsmath}
\usepackage{booktabs}
\shortauthors{Li et al.}

\newcommand{\lsun}{$L_{\sun}$}

\newcommand{\hto}{$\rm H_{2}O$}
\newcommand{\radpdr}{$G_{\rm 0}$}
\newcommand{\radxdr}{$F_{\rm x}$}
\newcommand{\nht}{$n(\rm H_{2}) $}
\newcommand{\htoo}{$\rm H_{2}O (3_{12}- 2_{21})$}
\newcommand{\htot}{$\rm H_{2}O (3_{21}- 3_{12})$}
\newcommand{\cotwo}{CO$(2-1)$}

\newcommand{\ci}{[\ion{C}{1}]$(2-1)$}
\newcommand{\oh}{$\rm OH^{+}(1_{1}-0_{1})$}    
\newcommand{\cii}{[\ion{C}{2}]$_{\rm 158 \mu m} $}
\newcommand{\cofive}{CO$(5-4) $}
\newcommand{\cosix}{CO$(6-5) $}
\newcommand{\coseven}{CO$(7-6) $}
\newcommand{\coeight}{CO$(8-7) $}
\newcommand{\conine}{CO$(9-8) $}
\newcommand{\coten}{CO$(10-9) $}
\newcommand{\kmps}{$\rm km\ s^{-1}$}

\newcommand{\msun}{$\rm M_{\sun}$}
\newcommand{\sfr}{$\rm M_{\sun}\ yr^{-1}$}
\newcommand{\qso}{P215$-$16}
\newcommand{\qsos}{J1429+5447}
\newcommand{\reds}{$z = 5.78$}
\newcommand{\redss}{$z = 6.18$}

\graphicspath{{./}{figures/}}

\begin{document}
\title{Diverse molecular gas excitations in quasar host galaxies at $z\sim 6$}

\correspondingauthor{Ran Wang}
\email{rwangkiaa@pku.edu.cn}

\correspondingauthor{Jianan Li}
\email{jiananl@mail.tsinghua.edu.cn}

\author[0000-0002-1815-4839 ]{Jianan Li}
\affiliation{Department of Astronomy, Tsinghua University, Beijing 100084, China}
\affiliation{Kavli Institute for Astronomy and Astrophysics, Peking University, Beijing 100871, China}

\author[0000-0003-4956-5742]{Ran Wang}
\affiliation{Kavli Institute for Astronomy and Astrophysics, Peking University, Beijing 100871, China}

\author[0000-0001-9815-4953]{Antonio Pensabene}
\affiliation{Dipartimento di Fisica “G. Occhialini”, Università degli Studi di Milano-Bicocca, Piazza della Scienza 3, I-20126, Milano, Italy}

\author[0000-0003-4793-7880]{Fabian Walter}
\affiliation{Max Planck Institut f\"ur Astronomie, K\"onigstuhl 17, D-69117, Heidelberg, Germany}
\affil{National Radio Astronomy Observatory, Pete V. Domenici Array Science Center, P.O. Box O, Socorro, NM 87801, USA}

\author[0000-0001-9024-8322]{Bram P. Venemans}
\affiliation{Leiden Observatory, Leiden University, PO Box 9513, 2300 RA Leiden, The Netherlands}

\author[0000-0002-2662-8803]{Roberto Decarli}
\affiliation{INAF -- Osservatorio di Astrofisica e Scienza dello Spazio, via Gobetti 93/3, 40129 Bologna, Italy}

\author[0000-0002-2931-7824]{Eduardo Ba\~nados}
\affiliation{Max Planck Institut f\"ur Astronomie, K\"onigstuhl 17, D-69117, Heidelberg, Germany}

\author[0000-0003-2027-8221]{Pierre Cox}
\affiliation{Sorbonne Universit\'e, UPMC Universit\'e Paris 6 and CNRS, UMR 7095, Institut d’Astrophysique de Paris, 98bis Boulevard Arago,75014 Paris, France}

\author[0000-0002-7176-4046]{Roberto Neri}
\affiliation{Institut de Radioastronomie Millimétrique (IRAM), 300 Rue de la Piscine, 38400 Saint-Martin-d’H\'eres, France}

\author[0000-0002-4721-3922]{Alain Omont}
\affiliation{Institut d'Astrophysique de Paris, Sorbonne Universit\'{e}, CNRS, UMR 7095, 98 bis bd Arago, 75014 Paris, France}

\author[0000-0001-8467-6478]{Zheng Cai}
\affiliation{Department of Astronomy, Tsinghua University, Beijing 100084, China}

\author[0000-0002-7220-397X]{Yana Khusanova}
\affiliation{Max Planck Institut f\"ur Astronomie, K\"onigstuhl 17, D-69117, Heidelberg, Germany}

\author[0000-0003-0754-9795]{Fuxiang Xu}
\affiliation{Department of Astronomy, School of Physics, Peking University, Beijing 100871,  China}
\affiliation{Kavli Institute for Astronomy and Astrophysics, Peking University, Beijing 100871, China}

\author[0000-0001-9585-1462]{Dominik Riechers}
\affiliation{I. Physikalisches Institut, Universit\"at zu K\"oln, Z\"ulpicher Strasse 77, D-50937 K\"oln, Germany}

\author[0000-0002-1815-4839 ]{Jeff wagg}
\affiliation{PIFI Visiting Scientist, Purple Mountain Observatory, No. 8 Yuanhua Road, Qixia District, Nanjing 210034, People's Republic of China}

\author[0000-0002-1478-2598]{Yali Shao}
\affiliation{School of Space and Environment, Beihang University, Beijing, China}

\author[0000-0001-9321-6000]{Yuanqi Liu}
\affiliation{Shanghai Astronomical Observatory, Chinese Academy of Sciences, 80 Nandan Road, Shanghai 200030, China}

\author[0000-0001-6459-0669]{Karl M. Menten}
\affiliation{Max-Planck-Institut f{\"u}r Radioastronomie, Auf dem H\"{u}gel 69, 53121 Bonn, Germany}

\author[0000-0002-3119-9003]{Qiong Li}
\affiliation{Jodrell Bank Centre for Astrophysics, University of Manchester, Oxford Road, Manchester M13 9PL, UK}

\author[0000-0003-3310-0131]{Xiaohui Fan}
\affiliation{Steward Observatory, University of Arizona, 933 N Cherry Avenue, Tucson, AZ 85721, USA}





\begin{abstract}
We present observations using the NOrthern Extended Millimetre Array (NOEMA) of CO and \hto{} emission lines, and the underlying dust continuum in two quasars at $z\sim 6$, i.e., \qso{} at \reds{} and \qsos{} at \redss{}.  Notably, among all published CO SLEDs of quasars at $z\sim 6$, the two systems reveal the highest and the lowest CO level of excitation, respectively. 
Our radiative transfer modeling of the CO SLED of \qso{} suggests that the molecular gas heated by AGN could be a plausible origin for the high CO excitation. 
For \qsos{}, we obtain the first well-sampled CO SLED (from transitions from 2-1 to 10-9) of a radio-loud quasar at $z\gtrsim 6$. 
Analysis of the CO SLED suggests that a single photo-dissociation region (PDR) component could explain the CO excitation in the radio-loud quasar \qsos{}. 
This work highlights the utility of the CO SLED in uncovering the ISM properties in these young quasar-starburst systems at the highest redshift. The diversity of the CO SLEDs reveals the complexities in gas conditions and excitation mechanisms at their early evolutionary stage.
\end{abstract}



\section{Introduction} \label{sec:intro}
Active star formation has been detected in the host galaxies of quasars from the local to the distant universe, revealing co-evolution of the supermassive black holes (SMBHs) and  galaxies (e.g., \citealt{kormendy13}; \citealt{wang13,wang16}; \citealt{willott15}; \citealt{venemans16}; \citealt{decarli17}; \citealt{pensabene20}). The cold ($\rm \sim 100\ K$) interstellar medium (ISM) in the quasar host galaxies fuels both the active galactic nucleus (AGN) and the nuclear star formation. Observations of the molecular and atomic lines at submillimeter/millimeter wavelengths provide rich information about the physical conditions and heating power of the cold ISM, which are the keys to understanding the evolution of the young quasar hosts (e.g., \citealt{wang11,wang13,wang16}; \citealt{decarli17,decarli18,decarli23}; \citealt{venemans18}; \citealt{shao19,shao22};  \citealt{li20,li20b};  \citealt{meyer22}).

The CO molecule, as the second most abundant species of the molecular ISM, gives the brightest molecular line emission in the (sub)millimeter regime. 
The CO spectral line energy distributions (SLEDs), i.e., the CO line fluxes as a function of rotational transition quantum number $J$, are sensitive diagnostics of the physical properties of the molecular gas (e.g., temperature, density, and radiation field strength). 
The low-$J$ ($J \leq 3$) CO emission lines directly constrain the total molecular gas mass of galaxies (e.g., \citealt{walter03}; \citealt{riechers06, riechers11a}; \citealt{wang11}; \citealt{shao19}; \citealt{ramos22}; \citealt{montoya23}).
Low-$J$ CO lines are found to be sublinearly correlated with far-infrared (FIR) luminosity (e.g., \citealt{riechers06, riechers11b}; \citealt{kamenetzky16}).
The mid-$J$ ($4 \leq J \leq 8$) CO lines are reported to be linearly correlated with FIR luminosity (e.g., \citealt{greve14}; \citealt{lu14}; \citealt{liu15}; \citealt{kamenetzky16}), which is consistent with the gas heated by the far-ultraviolet (FUV; 6 eV $< h\nu <$ 13.6 eV) photons from young-massive stars \citep{carilli13}. 
The high-$J$ ($J\geq 9$) CO lines require high temperature and density to be excited, and processes such as X-ray ($\sim 1-100$ keV) heating from the AGN, mechanical heating from shocks, and cosmic-ray heating are frequently proposed to explain their excitation (e.g., \citealt{weiss07}; \citealt{spinoglio12}; \citealt{meijerink13}; \citealt{li20}). Thus, the CO SLEDs also serve as probes for discriminating between different gas heating scenarios (\citealt{meijerink05,meijerink07}; \citealt{spaans08}), e.g., the photo-dissociation region (PDR), where the gas is heated by the far-ultraviolet (FUV) photons from young massive stars and the AGN-heated X-ray dominated region (XDR).

Searching for direct evidence of the influence of the AGN activity on the ISM excitation is observationally challenging even in local/low-$z$ well-studied samples. \cite{valentino21} find that for a sample of normal $z\sim 1 - 1.7$ galaxies that host AGNs, marginal correlations have been found between the presence and strength of AGNs and the CO excitation on galaxy scales. 
On nuclear scales of 250 pc, \cite{esposito22} report weak correlations between the CO excitation and either the X-ray or FUV flux in a sample of 35 local active galaxies, which suggests that neither the AGN nor star formation has a dominant effect on the gas excitation in the centers of these galaxies.
Studies of CO SLEDs of individual local AGN host galaxies suggest a mix of a PDR component that contributes to the low-to-mid $J$ CO fluxes and an XDR that dominates the high-$J$ CO emission (e.g., \citealt{van10}; \citealt{spinoglio12}; \citealt{pozzi17}; \citealt{mingozzi18}). 
Other mechanisms, such as mechanical heating by shocks, dense PDRs, and cosmic-ray heating, are 
also reported to dominate the high-$J$ CO excitation in some local AGN hosts and starburst galaxies (e.g., \citealt{van10}; \citealt{spinoglio12}; \citealt{pellegrini13}). 

Luminous quasars at high-$z$ represent the most luminous AGNs across redshifts. They are reported to be responsible for their high CO excitation at high-$J$. The lensed quasar APM 08279 at $z=3.9$, is currently the highest excited CO SLED ever published across redshifts. Its CO emission is detected within the central $\sim 550$ pc scale \citep{riechers09}, and the high CO excitation is best explained by the X-ray heating from the AGN (\citealt{weiss07}; \citealt{braford11}). Another example is the Cloverleaf quasar at $z=2.6$, for which the CO SLED is best fitted with a combination of a PDR and an XDR component \citep{uzgil16}.

Observations of quasars at $z\sim 6$ from optical to (sub-) millimeter wavelengths reveal that $\sim 1/3$ of them reside in extremely active environments with rapid SMBH accretion at rates of $\dot{M_{\rm BH}}$ $ > 10$ \sfr{} and intense star formation at rates up to a few $\sim$ 1000 \sfr{} in the host galaxy (e.g., \citealt{wang08}; \citealt{carilli13}; \citealt{derosa14};  \citealt{venemans20}).
In such systems, both the star formation and the SMBH activity lead to the injection of vast amounts of energy into the ISM of the host galaxy, making its ISM an ideal target to study the early co-evolution between the SMBHs and their host galaxies. 
Well-sampled CO SLEDs (with at least four transitions detected) have been obtained for five quasars at $z\sim 6$ (\citealt{bertoldi03}; \citealt{walter03}; \citealt{riechers09}; \citealt{gallerani14}; \citealt{carniani19}; \citealt{wangf19}; \citealt{yang19}; \citealt{li20}; \citealt{pensabene21}). Notably, all of them show highly-excited CO SLEDs with much brighter high-$J$ CO lines compared to that of the local starburst galaxies and AGNs, suggesting that the high-$J$ CO emission is likely dominated by an XDR component (\citealt{gallerani14}; \citealt{wangf19}; \citealt{yang19}; \citealt{li20}; \citealt{pensabene21}). 
However, the shapes of the CO SLEDs are diverse from object to object. For example, the CO SLED of J0100+2802 is double-peaked, suggesting more than one gas component \citep{wangf19}. In contrast, the CO SLED of J2310+1855 reveals one dominant gas component for the mid and high-$J$ CO transitions \citep{li20}.

In this paper, we use Northern Extended Millimeter Array (NOEMA) operated by the Institut de Radioastronomie Millim{\' e}trique  (IRAM) to sample the CO SLEDs of the quasar \qso{} at \reds{} and the quasar \qsos{} at \redss{}, each with at least five transitions ranging from \cofive{} to \coten{}. \qso{} is among the optically brightest quasars at $z\sim 6$, first identified in the Pan-STARRS1 survey \citep{morganson12}. It is also the source with the brightest 450 and 850 $\mu $m flux densities in our SCUBA2 High rEdshift bRight quasaR surveY (SHERRY) of $5.6<z<6.9$ quasars \citep{liq20}.   
\qsos{} is among the most distant radio-loud quasars at $z> 6$, discovered in the Canadian-French-Hawaii Quasar Survey (CFHQS; \citealt{willott10}). It is also among the most X-ray luminous quasar observed by eROSITA (\citealt{medvedev20}; \citealt{khorunzhev21}; \citealt{migliori23}), and the second (sub)millimeter brightest and the \cii{} brightest radio-loud quasar published at $z > 6$ \citep{khusanova22}. 

The structure of this paper is as follows. 
We describe observations and data reduction in Section \ref{observations} and observational results in Section \ref{results}. In Section \ref{fittingresults}, we derive the physical conditions of the dust and the gaseous ISM using dust spectral energy distribution models and  radiative transfer models, respectively. Discussions of the gas properties and comparison with other galaxy samples are presented in Section \ref{discussion}. We adopt a flat $\Lambda$CDM cosmology model with  $\rm H_{0} = 70\  km \ s^{-1} \ Mpc^{-1}$ and $\rm \Omega_{M}=0.3$, so 
that $1''$
 corresponds to 5.8 kpc and 5.6 kpc at \reds{} and \redss{}, respectively. 

\section{Observations and data reduction}\label{observations}
We obtained NOEMA 3 and 2 mm observations of the quasars \qso{} at \reds{} and \qsos{} at \redss{}, targeting the \cofive{}, \cosix{}, \coseven{}, \coeight{}, \conine{}, \coten{}, \oh{}, \ci{}, \htoo{}, and \htot{} emission lines, together with the underlying dust continuum emission.  The observations were carried out in three separate projects (programme ID: W0C3, W18EE, and S20CW), starting on 24th December 2012, and finishing on 27th December 2020, a period where number of antennas increased from 6 to 11. 

We used three separate executions in the compact  C/D array configurations to observe the \cofive{}, \cosix{},  \coeight{}, \conine{}, \coten{}, \oh{}, \htoo{}, and \htot{} emission lines, as well as the dust continuum emission of the quasar \qso{}.
Observations of the \cofive{} and \cosix{} emission lines were performed with an on-source time of 2.3 hours. 
We tuned the PolyFix correlator to cover the \cofive{} line in the lower sideband (LSB), and the \cosix{} line in the upper sideband (USB), each with a 7.744 GHz bandwidth. 
The \coeight{}, \conine{} and \oh{} lines were observed in one frequency setup for 1.9 hours on source, with the \coeight{} and \conine{}/\oh{} lines observed in the LSB and USB, respectively. 
We simultaneously observed the \conine{}, \coten{}, \oh{}, \htoo{}, and \htot{} lines with one tuning for 1.7 hours on source, with the \conine{} and \oh{} lines covered in the LSB and the \coten{}, \htoo{}, and \htot{} lines covered in the USB. 
The flux calibrator was MWC349 and the phase/amplitude calibrators were 1334-127, 1437-153, 1406-076, and 1435-218.

\begin{figure*}
\includegraphics[width=0.5\textwidth]{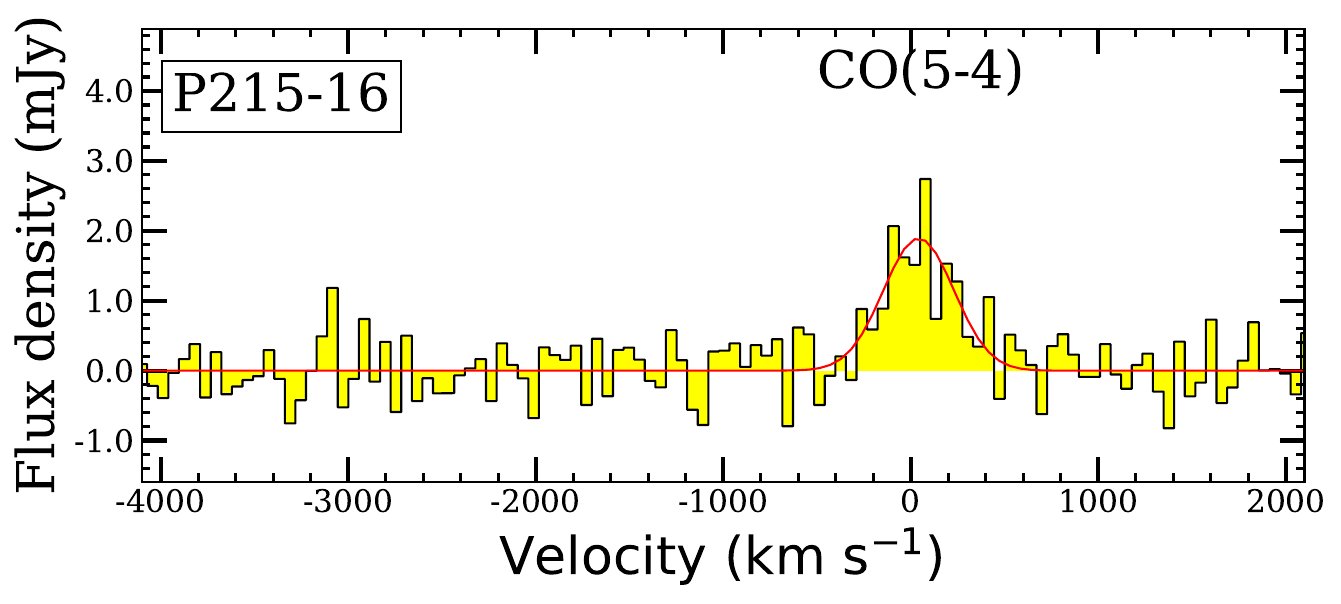}
\includegraphics[width=0.5\textwidth]{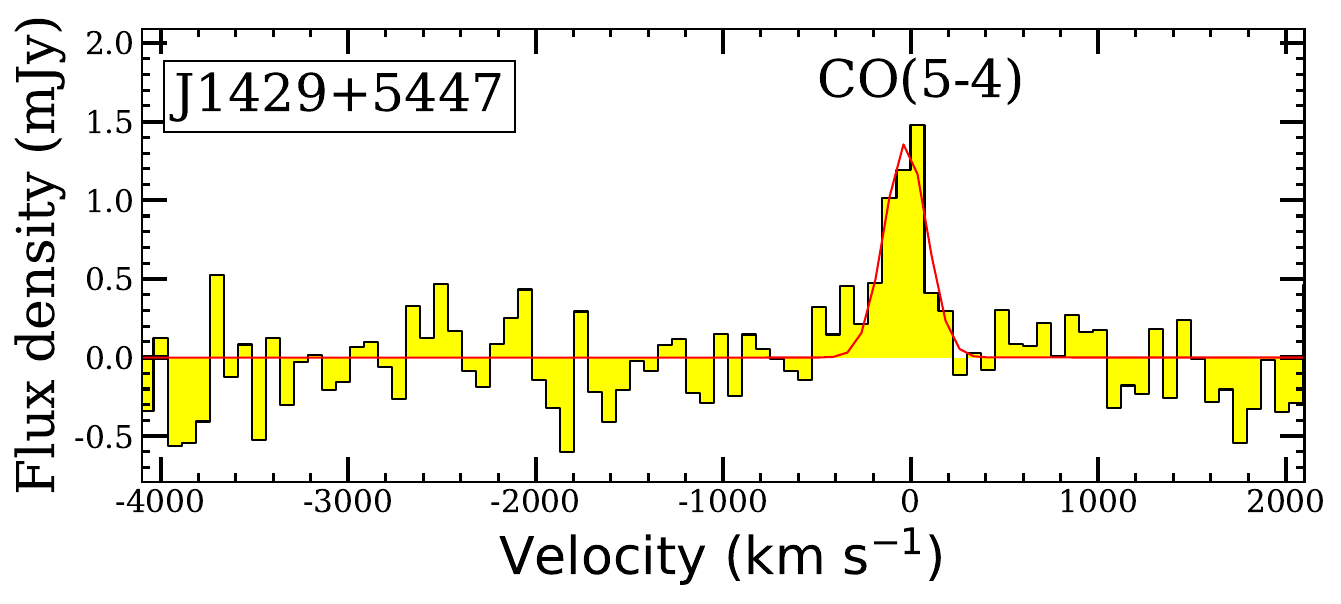}
\includegraphics[width=0.5\textwidth]{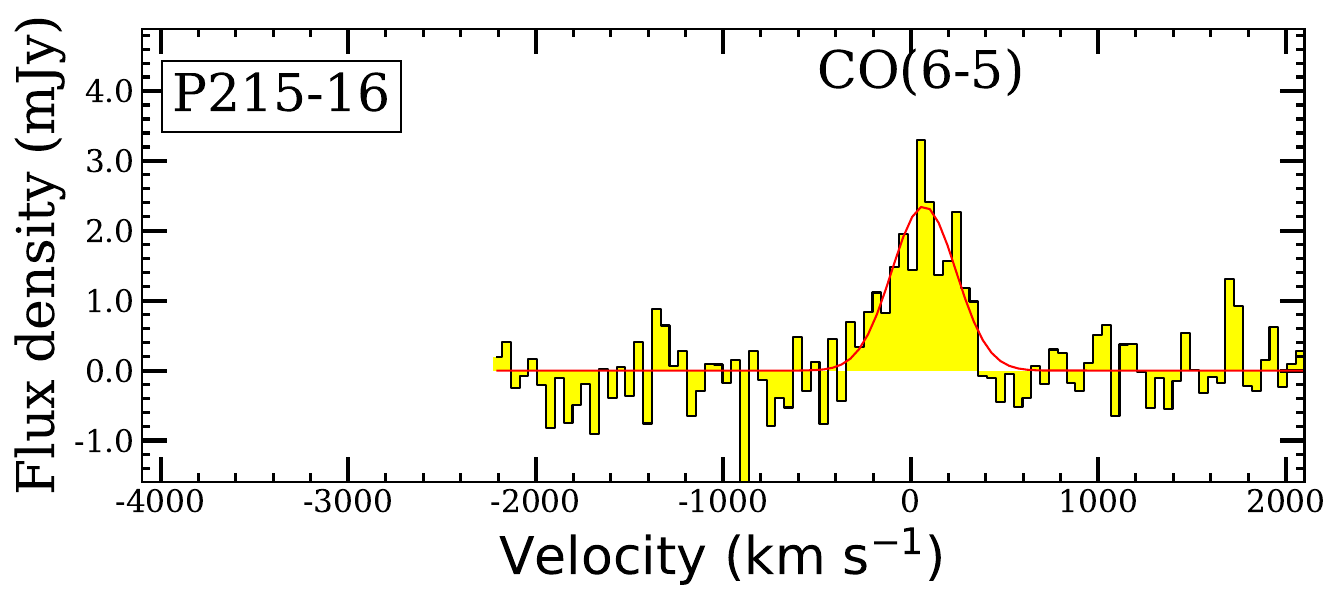}
\includegraphics[width=0.5\textwidth]{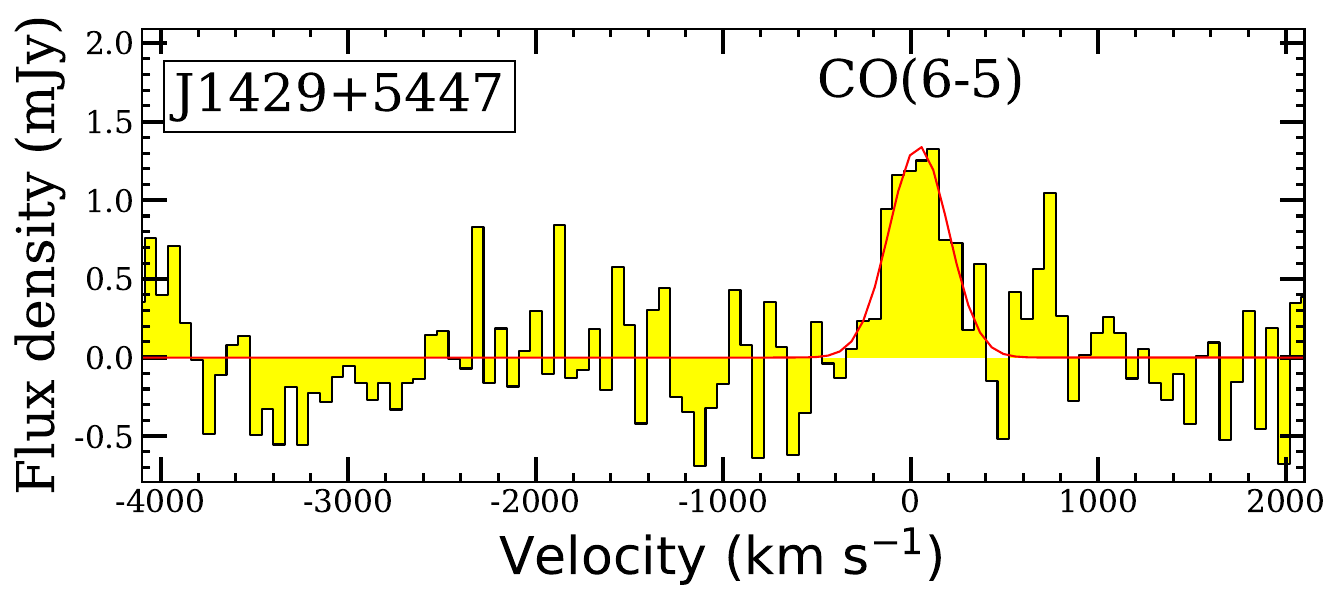}
\includegraphics[width=0.5\textwidth]{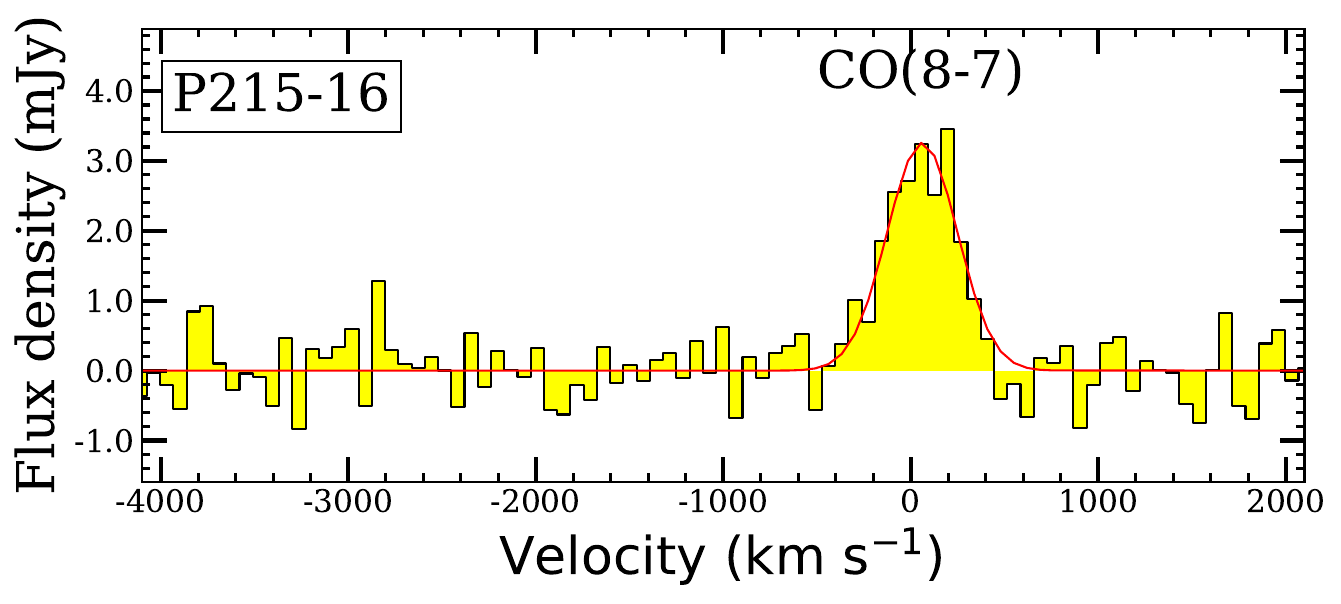}
\includegraphics[width=0.5\textwidth]{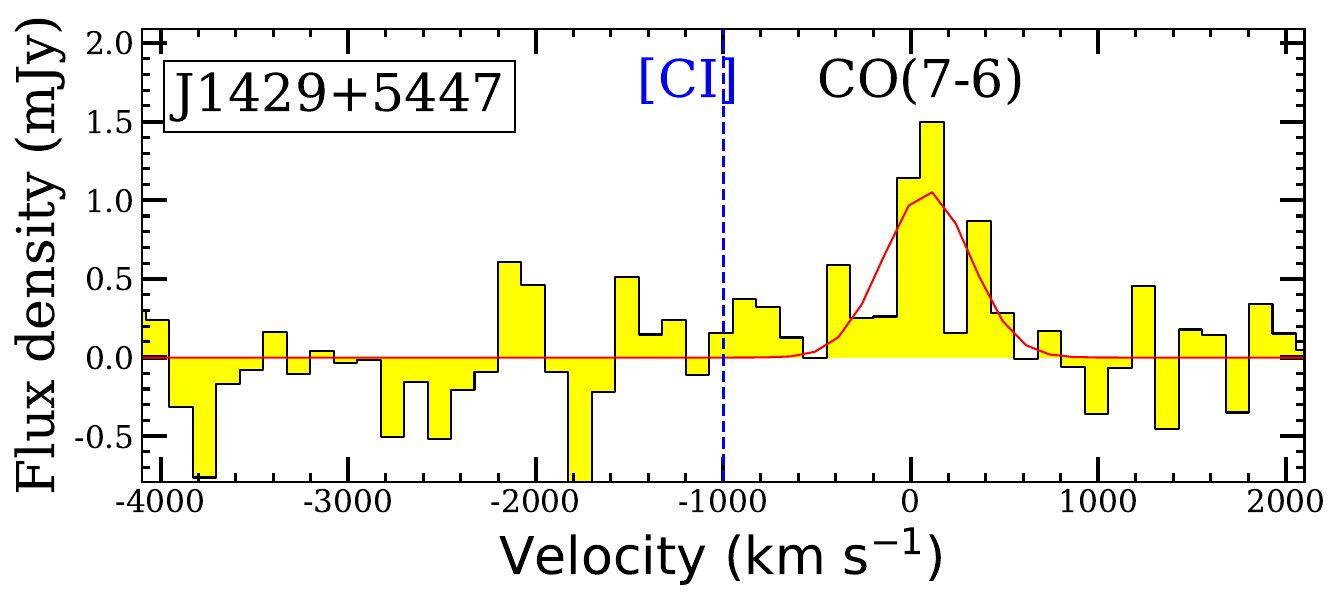}
\includegraphics[width=0.5\textwidth]{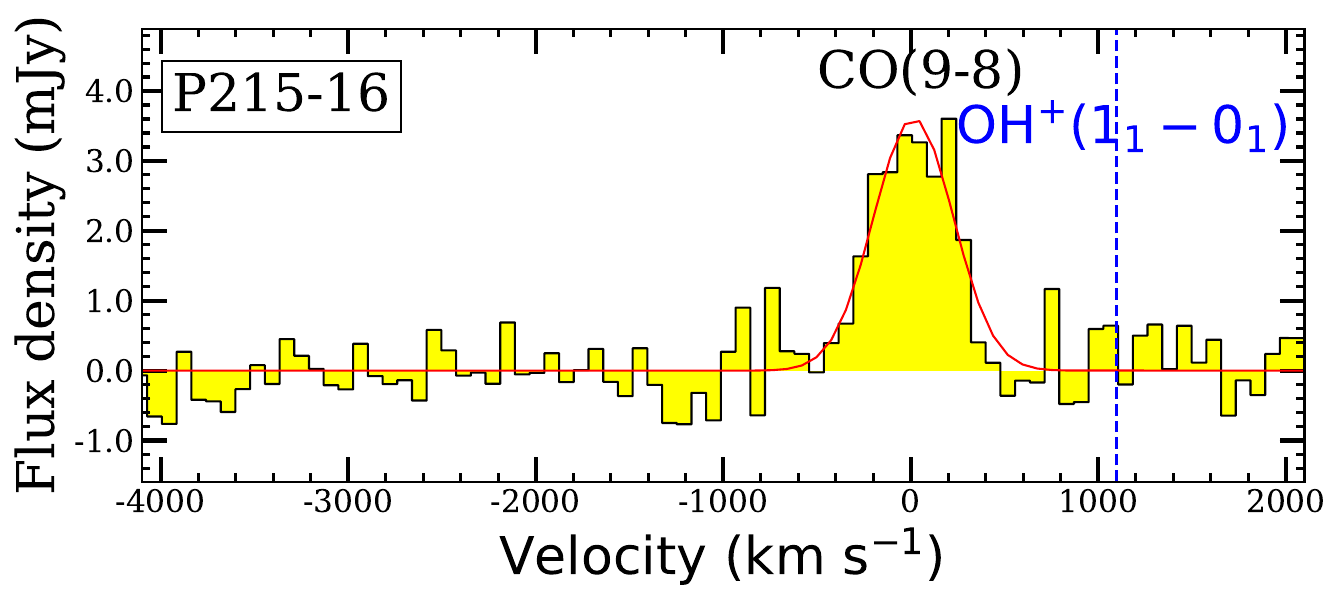}
\includegraphics[width=0.5\textwidth]{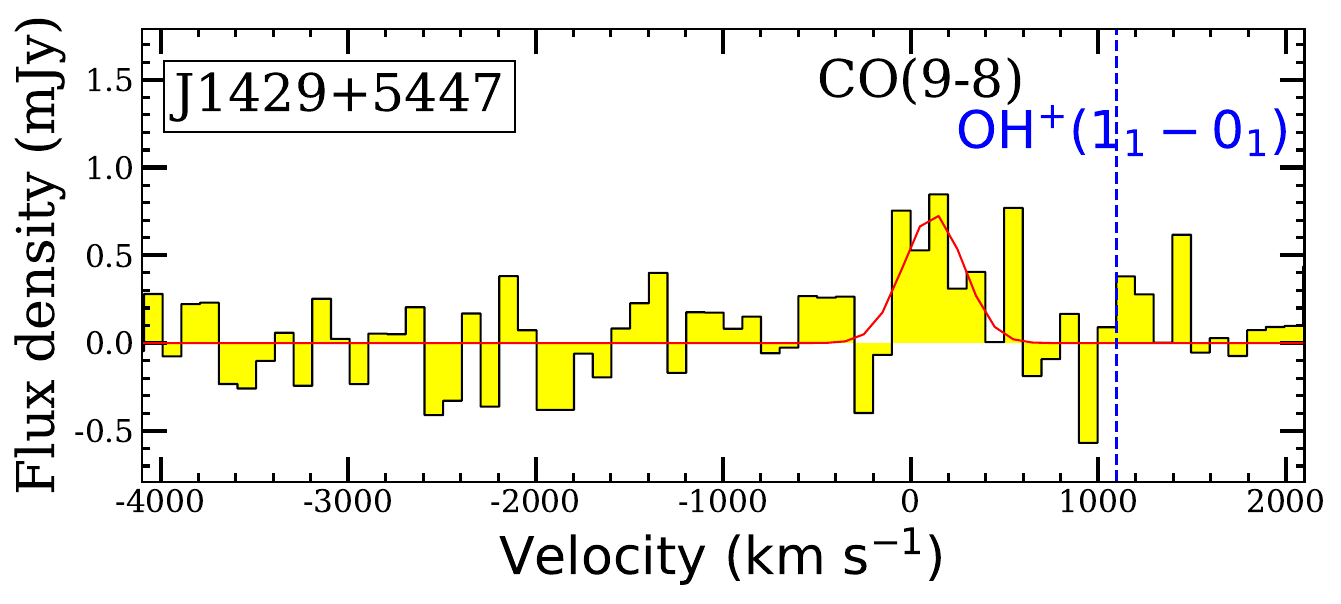}
\includegraphics[width=0.5\textwidth]{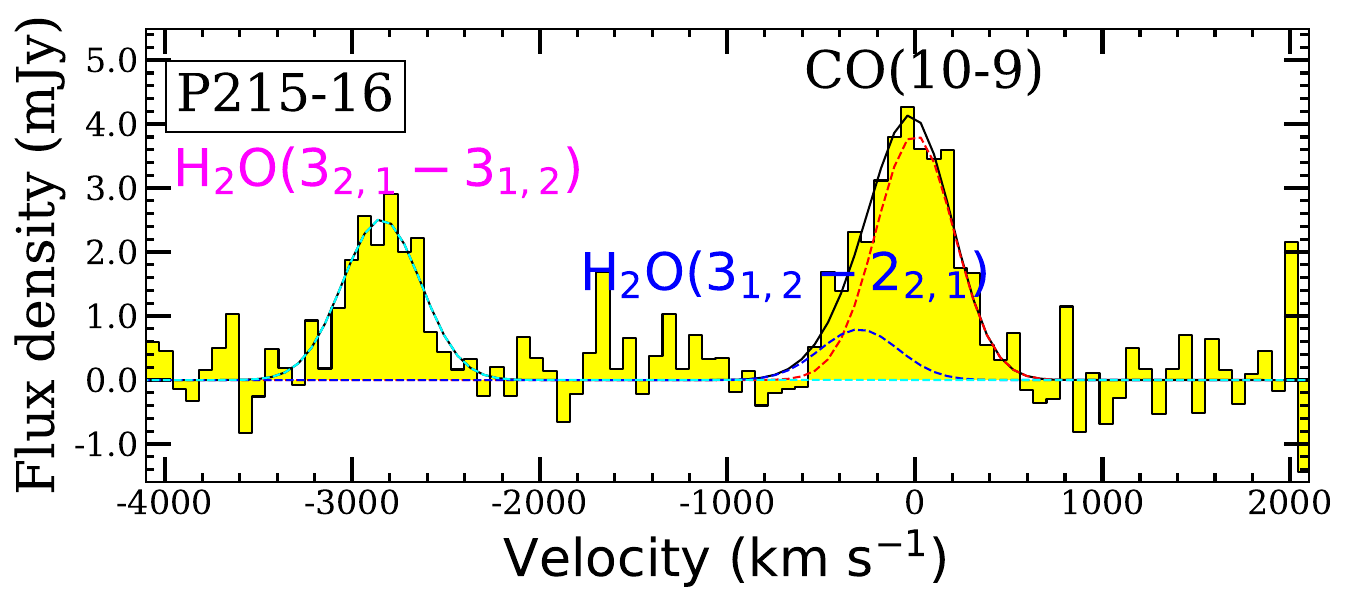}
\includegraphics[width=0.5\textwidth]{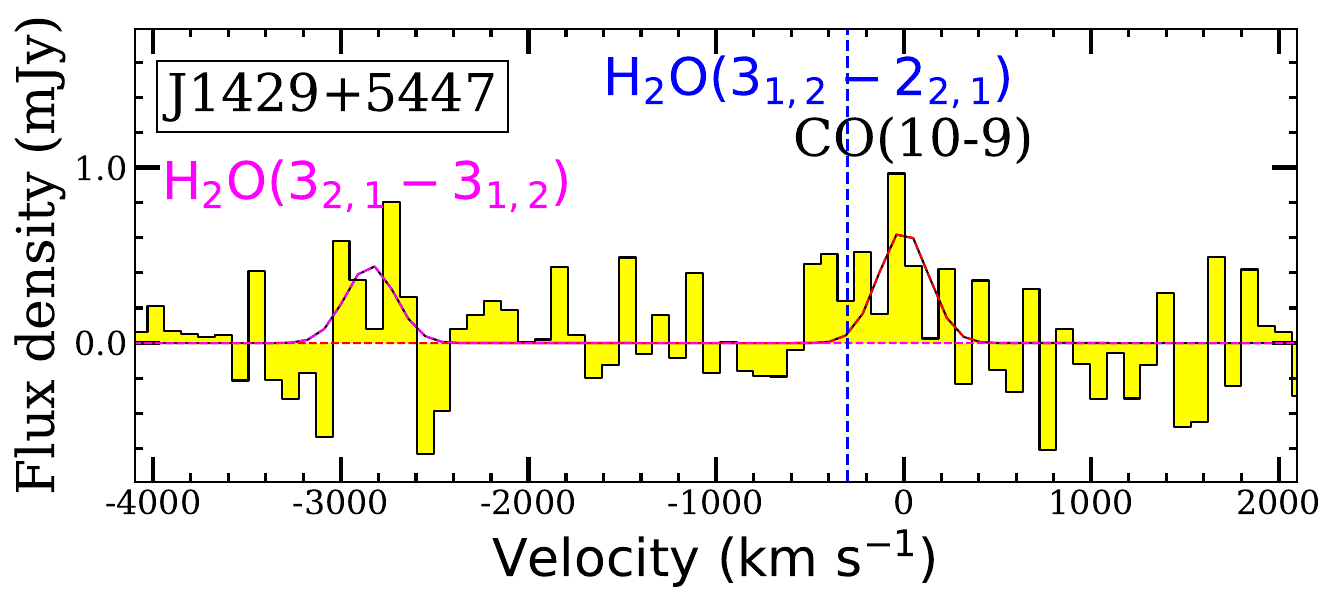}
\caption{Spectra of the CO, \hto{}, \oh{}, \ci{} lines for \qso{} (left) and \qsos{} (right). The continuum  emission has been subtracted for each of the spectra. 
The data are shown as yellow histograms. 
Gaussian profiles fitted to the line emission are shown as solid or dashed lines.
The expected positions of the undetected lines are indicated as dashed blue vertical lines. 
In the spectra where more than one line is detected, the solid black lines represent the sum of different Gaussian components. }
\label{fig1} 
\end{figure*}

As for the quasar \qsos{}, we used three separate executions to cover the \cofive{}, \cosix{}, \coseven{},\ci{}, \conine{}, \oh{}, \coten{}, \htoo{}, \htot{} emission lines, and the underlying continuum emission.
The \cofive{} and the \cosix{} lines were covered in the LSB and USB, respectively, with one frequency setup, for an on-source time of 3.2 hours.  The \conine{}, \oh{}, \coten{}, \htoo{}, and the \htot{} emission lines were observed simultaneously in one tuning in the 2 mm band for 2.1 hours on source. 
Observations of the \coseven{} and \ci{} lines were conducted in one frequency setup for 2.2 hours on source. 
We observed the continuum emission using all line-free channels in the USBs and LSBs in each observation. 
For all the observations, 1418+546 was used as phase/amplitude calibrator. The fluxes were calibrated with MWC349, 2010+723 and 3C273. 
The typical calibration uncertainty is $<10 \%$ in Band 1 and $<15 \%$ in Band 2.

We reduced the data using the CLIC and MAPPING packages that are part of the Grenoble Image and Line Data Analysis System (GILDAS) software  \citep{guilloteau2000}. We obtained the continuum $uv$ tables by averaging all the line-free channels using the UV$
\_$AVERAGE task in MAPPING. To generate the continuum subtracted spectral line $uv$ tables, we adopted the continuum $uv$ table obtained above  as a continuum model, and use the UV$\_$SUBTRACT task to subtract it from the original $uv$ tables. 
The $uv$ tables of the spectral lines and the continuum were cleaned and imaged with natural weighting to maximize the S/N ratio (SNR) for point source. 
The host galaxies are spatially unresolved in beam sizes of $\sim$ 3$^{\prime\prime}$--6$^{\prime\prime} \times 1^{\prime\prime}$--2$^{\prime\prime}$ for \qso{} and 2$^{\prime\prime}$--7$^{\prime\prime} \times 2^{\prime\prime}$--5$^{\prime\prime}$ for \qsos{}, respectively. 
The spectra were finally binned to 46--78 \kmps{} for the quasar \qso{} and 45--75 \kmps{} for the quasar \qsos{}. The rms noise for the continuum emission in all bands is 0.02--0.05 mJy $\rm beam^{-1}$ for both sources.
Detailed information on the targeted lines and derived parameters can be found in Table \ref{tabline}.

\begin{deluxetable*}{ccccccccccccc}[ht] 
\tabletypesize{\tiny}
\tablecaption{Measured emission line and continuum properties \label{tabline}}
\tablecolumns{13}
\tablehead{
\colhead{Source}   &\colhead{Line} &$\nu_{rest}$&\colhead{$z$}  &  \colhead{FWHM} & \colhead{S$\delta v$}   
&\colhead{Luminosity}& \colhead{Beam Size} &\colhead{Channel width} \\
 \colhead{}  & \colhead{}  &GHz & &  \colhead{(km s$^{-1}$)} & \colhead{(Jy $\rm km\ s^{-1}$)}   & \colhead{($\times 10^{8}\ L_{\sun}$)}&&\colhead{($\rm km\ s^{-1}$)}\\
\colhead{(1)}&\colhead{(2)}&\colhead{(3)}&\colhead{(4)}&\colhead{(5)} &\colhead{(6)}&\colhead{(7)} &\colhead{(8)}&\colhead{(9)} 
}
\startdata
\qso{}&\cofive{}&576.268&5.7831 $\pm$ 0.0006&444 $\pm$ 63&0.90 $\pm$ 0.17& 2.4 $\pm$ 0.5&5\farcs{74}$\times$1\farcs{73}, PA=5\textdegree &56\\
&\cosix{}&691.473&5.7837 $\pm$ 0.0004&404 $\pm$ 42&1.01 $\pm$ 0.14&3.3 $\pm$ 0.5&4\farcs{82}$\times$1\farcs{45}, PA=4\textdegree &46\\
&\coeight{}&921.800&5.7834 $\pm$ 0.0003&443 $\pm$ 39&1.54 $\pm$ 0.18&6.7 $\pm$ 0.8&  4\farcs{45}$\times$0\farcs{92}, PA=8\textdegree&70\\
& \oh{} &1033.118&5.7824&503&$<$ 0.57&$<$ 2.7&3\farcs{14}$\times$1\farcs{09}, PA=5\textdegree &78\\
&\conine{}&1036.912&5.7824 $\pm$ 0.0004&503 $\pm$ 37&1.93 $\pm$ 0.19&9.4 $\pm$ 0.9&3\farcs{14}$\times$1\farcs{09}, PA=5\textdegree &78\\
&\coten{}&1151.985&5.7824&491$\pm$41& 2.00 $\pm$ 0.24&10.8 $\pm$ 3.6&4\farcs{41}$\times$1\farcs{09}, PA=6\textdegree&71\\
&\htoo{}&1153.127&5.7824&491$\pm$41&0.42 $\pm$ 0.18&2.2 $\pm$ 0.9&4\farcs{41}$\times$1\farcs{09}, PA=6\textdegree&71\\
&\htot{}&1162.912&5.7824&491$\pm$41&1.32 $\pm$ 0.18&7.2 $\pm$ 2.0&4\farcs{41}$\times$1\farcs{09}, PA=6\textdegree&71\\
\hline    
\qsos{}&\cofive{}&576.268&6.1824 $\pm$ 0.0005&267 $\pm$ 50&0.39 $\pm$ 0.09&1.2 $\pm$ 0.3&6\farcs{32}$\times$5\farcs{11}, PA=46\textdegree&75\\
&\cosix{}& 691.473&6.1841 $\pm$ 0.0007&372 $\pm$ 70&0.53 $\pm$ 0.13&1.9 $\pm$ 0.5&5\farcs{33}$\times$4\farcs{33}, PA=48\textdegree&62\\
&\coseven{}&806.652&6.1841&372&0.48 $\pm$ 0.10&2.0 $\pm$ 0.4&2\farcs{45}$\times$2\farcs{14}, PA=38\textdegree&62\\
&\ci{}&809.342&6.1841&372&$<0.31$&$<1.3$&2\farcs{45}$\times$2\farcs{14}, PA=38\textdegree&62\\
&\oh{} &1033.118&6.1860&271&$<$ 0.21&$<$ 1.1&3\farcs{85}$\times2$\farcs{71}, PA=51\textdegree&50\\
&\conine{}&1036.912&6.1860 $\pm$ 0.0012&$271 \pm 69$&0.29 $\pm$ 0.07&1.3 $\pm$ 0.4&3\farcs{85}$\times2$\farcs{71}, PA=51\textdegree&50\\
&\coten{}&1151.985&6.1860&313 $\pm$ 119&0.21 $\pm$ 0.06&1.2 $\pm$ 0.3& 3\farcs{49}$\times2$\farcs{44}, PA=51\textdegree   &45\\
&\htoo{} &1153.127&6.1860&313&$<$ 0.18&$<$ 1.1& 3\farcs{49}$\times2$\farcs{44}, PA=51\textdegree & 45\\
&\htot{}&1162.912&6.1860&313 $\pm$ 119&$0.15 \pm 0.06$&$0.9 \pm 0.3$& 3\farcs{49}$\times2$\farcs{44}, PA=51\textdegree&45\\
\hline
& \multicolumn{7}{c}{\bf{Dust continuum properties}} \\
& \multicolumn{2}{c}{\qso{}} & & \multicolumn{5}{c}{\qsos{}} \\
\cmidrule(lr){1-4}\cmidrule(lr){6-9}
  $\rm \nu_{cont}$ & Flux density & Rms &Ref & & $\rm \nu_{cont}$ & Flux density & Rms& Ref\\
GHz&mJy& $\mu$Jy beam$^{-1}$ &&&GHz&mJy& $\mu$Jy beam$^{-1}$\\
(10)&(11)&(12)&(13)&&(14)&(15)&(16)&(17)\\
83.2 & 0.21 $\pm$ 0.02&21 &[1] &&82.2& 0.15  $\pm$ 0.02&19& [1] \\
98.6 &0.27 $\pm$ 0.02&20&[1]&&98.4 &0.09 $\pm$ 0.02 &20& [1]\\
138.3 &1.38 $\pm$ 0.03&29 &[1]&&112.4 &0.34 $\pm$ 0.05 &52& [1]\\
153.2 &1.84 $\pm$ 0.03&33&[1]&& 144.2 &0.32 $\pm$ 0.02&23& [1]\\  
168.4 &2.37 $\pm$ 0.05&48&[1]&&159.8 &0.41 $\pm$ 0.04&35& [1]\\
352.7 &16.85 $\pm$ 1.10 &\nodata&[2]&&1.4 &2.68 $\pm$ 0.16&\nodata&[3]\\ 
666.2 &26.93 $\pm$ 7.78& \nodata&[2]&&3 &1.81 $\pm$ 0.11&\nodata&[4]\\ 
&&&&& 32 &0.26 $\pm$ 0.02&\nodata&[5]\\
&&&&&250 &3.05 $\pm$ 0.11&\nodata&[6]\\
\enddata
\tablecomments{Columns 1--2: Source name and line ID. Columns 3--7: Line rest frequency, Redshift, line width in FWHM, flux, and luminosity. Columns 8--9: Beam size and channel width. Columns 10--13 and 14--17: Continuum frequency/wavelength, flux, RMS noise, and references for \qso{} and \qsos{}, respectively. References: [1] This work, [2] \citealt{liq20}, [3] \citealt{villarreal18}, [4] \citealt{becker95}, [5] \citealt{wang11}
, [6] \citealt{khusanova22}.}
\end{deluxetable*}

\section{results}\label{results}
We detect the \cofive{}, \cosix{}, \coeight{}, \conine{}, \coten{}, and the \htot{} lines from the host galaxy of \qso{} at SNR $>5$, and report a non-detection of the \oh{} line (Fig. \ref{fig1}). A tentative detection ($\sim2.3 \sigma$) of the \htoo{} line blended with the \coten{} line is also obtained for this object. The continuum emission is detected at SNR ratio $>10$ in all the observed bands (Table \ref{tabline}).
We extract the spectra from the peak pixels and fit the spectra of \cofive{}, \cosix{}, \coeight{}, and \conine{} lines separately, each with a Gaussian profile. 
The redshifts and line widths derived for different CO lines are consistent with each other within the uncertainties. 
The  \coten{}, \htoo{}, and \htot{} lines are covered in the same sideband. We fit these three lines simultaneously, using a Gaussian model,  with the line centers fixed to the redshift derived from the \conine{} line and an identical line width for all of the three spectral lines (assuming that these three lines are originating from the same gas component).
The spectra and the Gaussian fitting results are shown in Fig. \ref{fig1}. 
Adopting the line width of the \conine{} line, we obtain a velocity-integrated map centered at the \oh{} line frequency and measure the 3$\sigma$ detection upper limit for the \oh{} line.  
The measurements for both spectral lines and dust continuum are presented in Table \ref{tabline}.   

For the quasar \qsos{}, we detect the \cofive{}, \cosix{}, \coseven{}, and \conine{} lines at SNR $>4$, and tentatively detect the \coten{} and the \htot{} lines at the 3.5 and 2.5$\sigma$ levels, respectively. The \ci{}, \htoo{}, and \oh{} lines are undetected (Fig. \ref{fig1}). 
The CO lines detected in \qsos{} has lower S/N compared to those detected in \qso{}.
The SNR of the \coseven{} line spectrum is low.  We fix the Gaussian line width and redshift to that of the \cosix{} line and fit the line flux. 
The resulting line widths, redshifts, fluxes, and continuum fluxes are listed in Table \ref{tabline}. 
The $3 \sigma$ upper limits of the \ci{}, \oh{}, and \htoo{} lines are estimated based on the rms of the moment 0 maps, which are integrated over channels expected for these lines assuming the same line widths and redshifts as that of their neighboring CO lines. 
The derived redshifts and line widths of the CO lines are consistent with each other within the uncertainties. These are also consistent with the fitting results of the W component of the \cotwo{} line (the western peak that coincides with the optical quasar) reported in \citet{wang11}.  The continuum emissions are all detected at a $>4 \sigma$ level. More details about the measured dust and spectral line properties are listed in Table \ref{tabline}.

\section{dust and gas properties} \label{fittingresults}

\subsection{Dust SED fitting}\label{dustsed}

To derive the dust properties of the two quasars, we analyze their dust spectral energy distributions (SEDs).
Optically thick dust model has been suggested to explain the dust emission in a few high-$z$ starburst systems, especially at rest-frame wavelengths $< \sim 200\ \mu $m (e.g., \citealt{riechers13}; \citealt{wangf19}). 
Since the quasars in this work are all far-infrared bright sources,  it is reasonable to consider that the dust can be optically thick close to the peak of the dust SED. 
We thus employ the general opacity dust formula published in \cite{dacunha21} to fit the dust SEDs of the quasars in this work. In the general opacity regime, the dust emission can be described by:
\begin{align}
S_{\nu} \propto [1- {\rm exp(-\tau_{\nu})}] (B_{\nu}[T_{\rm dust}(z)]-B_{\nu}[T_{\rm CMB}(z)]),
\end{align}
taking into account the effect of the cosmic microwave background (CMB).
Here $B_{\nu}[T_{\rm dust}(z)]$ or $B_{\nu}[T_{\rm CMB}(z)]$ is the Planck function at the temperature of the dust or CMB at the quasar's redshift, and the dust optical depth $\tau_{\nu}$ is proportional to the dust surface density $\Sigma_{\rm dust}$ via:
\begin{align}
\tau_{\nu} = \kappa_{\nu} \Sigma_{\rm dust},
\end{align}
where $\kappa_{\nu}$ is the dust opacity. Its dependence on frequency is:
\begin{align}
\kappa_{\nu} = \kappa_{0} (\frac{\nu}{\nu_{0}})^{\beta},
\end{align}
where $\beta$ is the dust emissivity index and $\kappa_{0}$ is the emissivity of dust grains per unit mass at frequency $\nu_{0}$. If we assume a simple spherical geometry, then the dust mass is:
\begin{align}
M_{\rm dust} = 4\pi R^{2} \Sigma_{\rm dust},
\end{align}
where $R$ is the dust radius of a galaxy.
In \cite{dacunha21}, an additional parameter $\lambda_{\rm thick}$ is introduced, which is defined when the optical depth reaches 1,
\begin{align}
\tau_{\lambda_{\rm thick}} = 1 \Longrightarrow \lambda_{\rm thick} = \lambda_{0}(\kappa_{0}\frac{M_{\rm dust}}{4\pi R^{2}})^{1/\beta}.\label{eq5}
\end{align}
Accordingly, the dust emission can be rewritten as:
\begin{align}
S_{\nu} \propto \ & [1- {\rm exp(-(\frac{\nu \lambda_{thick}}{\textit{c}})^{\beta})}] \times \label{eq6}  && \\ 
& (B_{\nu}[T_{\rm dust}(z)]-B_{\nu}[T_{\rm CMB}(z)]),  && \nonumber
\end{align}
where $c$ is the speed of light. 

Complemented with SCUBA 2 observations of the dust continuum at 450 $\mu $m (666.2 GHz) and 850 $\mu $m (352.7 GHz) from \cite{liq20}, we use the general opacity model (Eq. \ref{eq6}) to fit the observed dust SED of \qso{}, normalized to the continuum flux density at 666.2 GHz. Under the general opacity assumption, the dust SED is determined by three parameters: namely $T_{\rm dust}$, $\beta$, and $\lambda_{\rm thick}$.
We consider three different dust models. 
To break the degeneracy between different parameters in dust SED fitting, we fix $\lambda_{\rm thick}$ to (1) the average value of  SMGs ($\lambda_{\rm thick}=100\ \mu m$; \citealt{dacunha21}) (Model 1), and (2) the average value of QSOs at $z\sim 6$ ($\lambda_{\rm thick}=264\ \mu m$; derived from \citealt{gilli22}) (Model 2), while enabling $T_{\rm dust}$ and $\beta$ to vary in these two models. 
In a third model, we enable all three parameters $\lambda_{\rm thick}$, $T_{\rm dust}$, and $\beta$ to vary (Model 3).
To explore the posterior distributions of the parameters, we use the Markov Chain Monte Carlo (MCMC) python package emcee to fit the data to the model \citep{foreman13}, adopting the best-fit values obtained for the least square estimation as educated initial guesses of the parameters. 

\begin{figure*}
\gridline{\fig{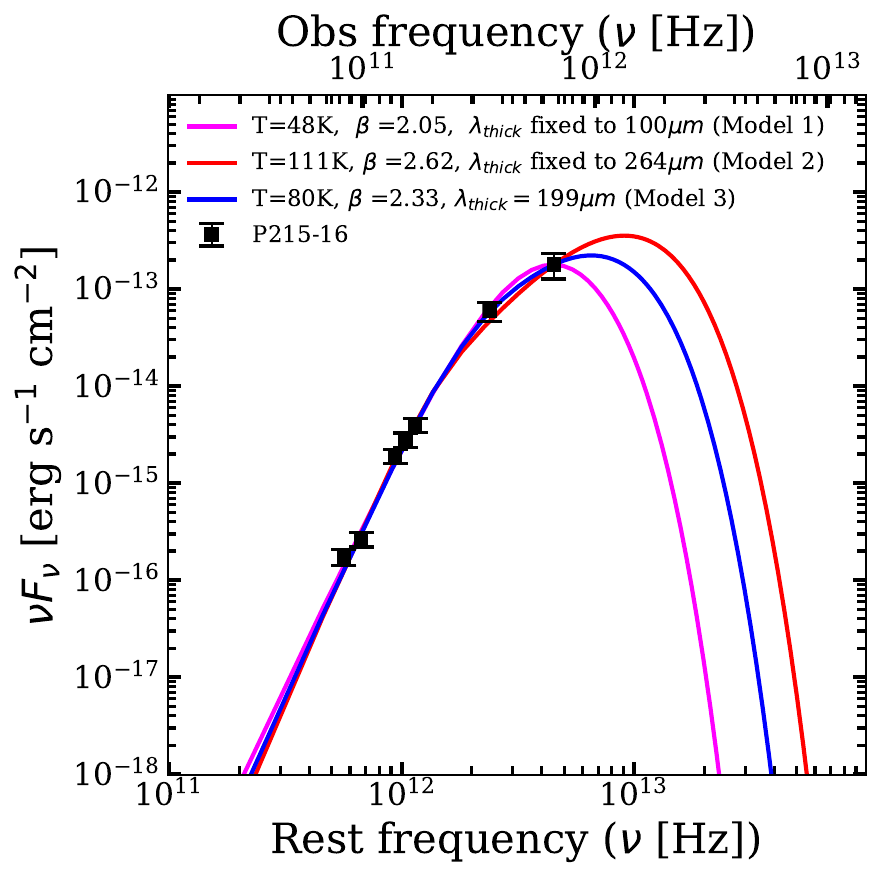}{0.5\textwidth}{(a)}
          \fig{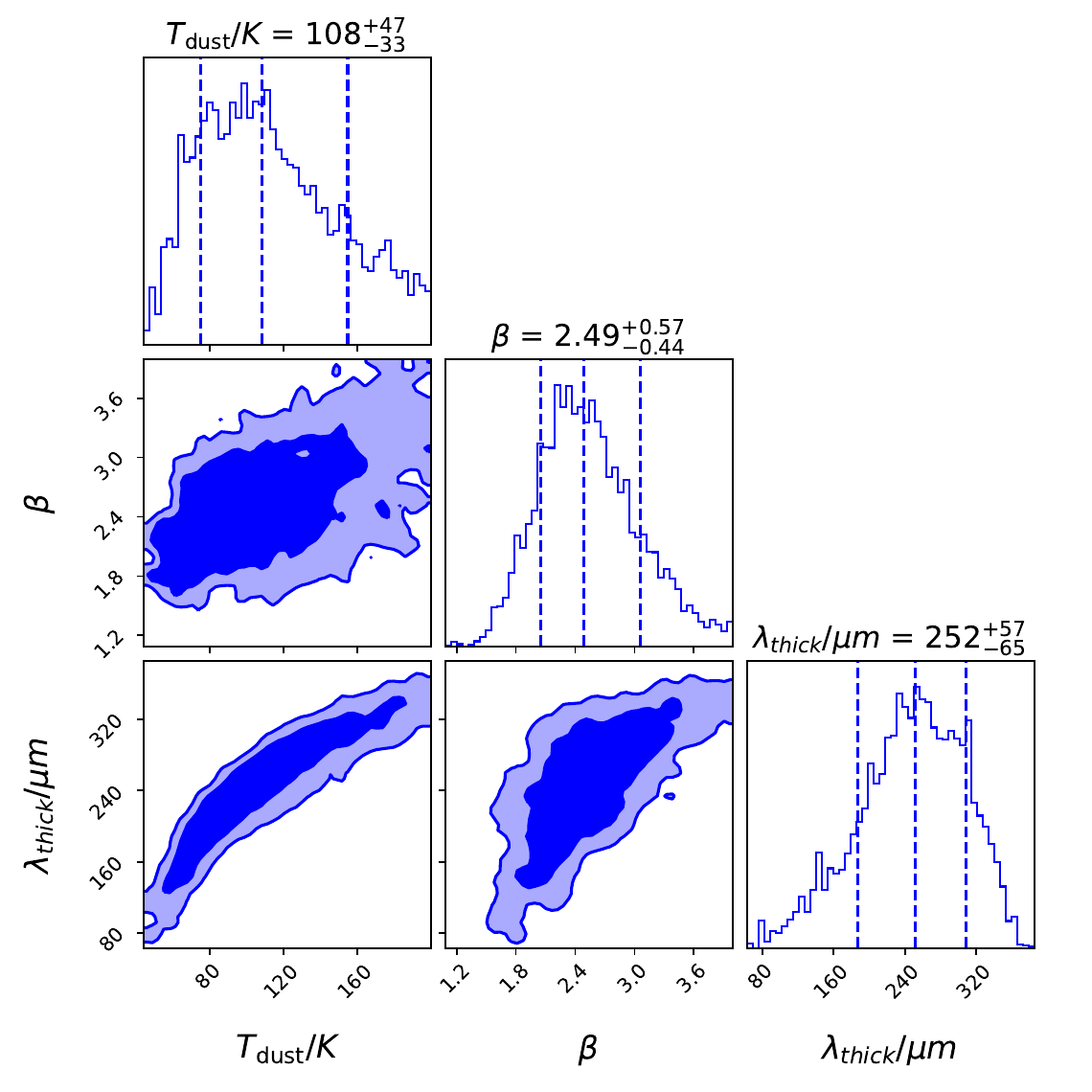}{0.5\textwidth}{Model 3}
          }
\gridline{\fig{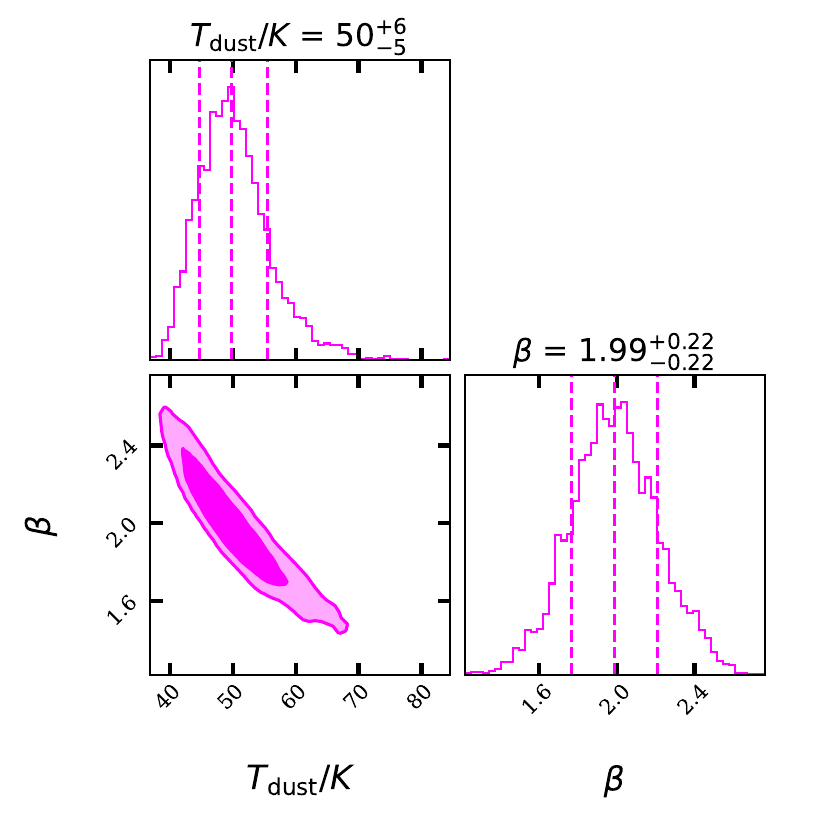}{0.5\textwidth}{Model 1 }
          \fig{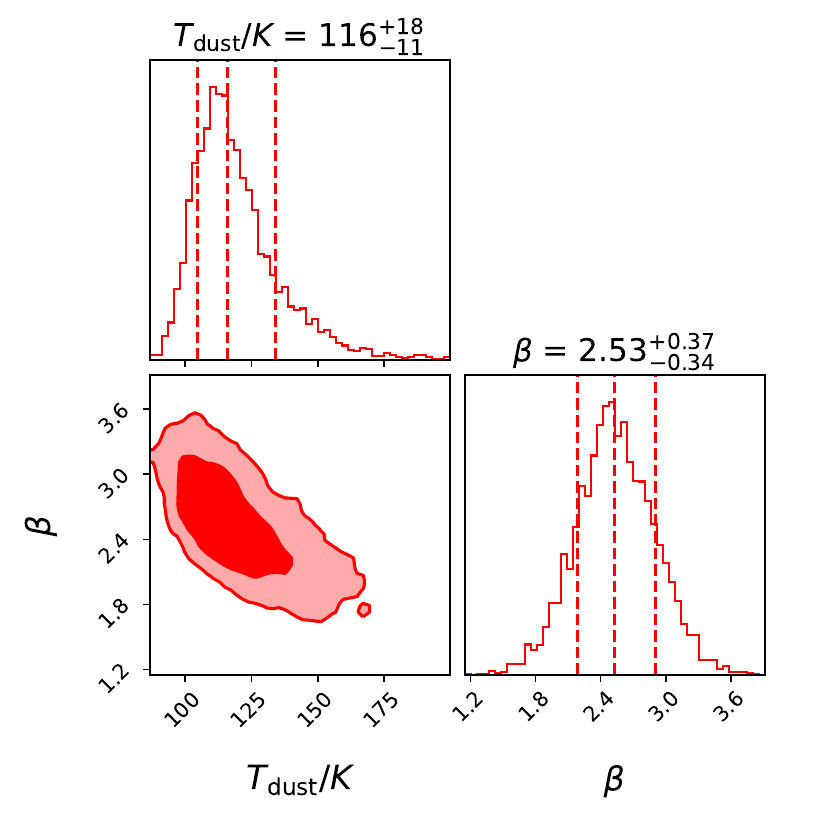}{0.5\textwidth}{Model 2 }
          }
\caption{\label{psosed1} (a) Dust SED fitting results for \qso{}. The data are shown as black squares. The Top and bottom axes show the observed and rest continuum frequency. The dust SED is fitted with three general opacity models, and the solid lines represent the models with best-fit values (see Sect. \ref{dustsed} for further details). 
(b--d) Posterior distributions of the model parameters for Model 3, 1 and 2, with the same color as the best-fit models in (a). The 2D contours show the [1, 2]$\sigma$ confidence intervals in the marginalized distributions. }
\end{figure*}

The inclusion of the high-frequency SCUBA 2 observations is critical in constraining the dust SED parameters for \qso{}. 
The least-square estimation finds best-fit values of  $T_{\rm dust} = 48 \rm \ K$, and $\beta = 2.05$ for Model 1,  $T_{\rm dust} = 111 \rm \ K$, and $\beta = 2.62$ for Model 2, and  $T_{\rm dust} = 80 \rm \ K$, $\beta = 2.33$, and $\lambda_{\rm thick} = 199\ \mu m$ for Model 3. 
In some studies, $T_{\rm dust}$, $\beta$ and $\tau_{1900GHz} $ are adopted to describe the general opacity dust SED model (e.g., \citealt{walter22}; \citealt{decarli23}), where $\tau_{1900GHz} $ is the dust optical depth at 1900 GHz.
The derived $\lambda_{\rm thick}$ values for our model translate into $\tau_{1900GHz} $ of 0.39, 3.84, and 1.71 for Model 1, 2, and 3, respectively, which is consistent with values derived for other quasars at $z\sim 6$ (\citealt{walter22}; \citealt{decarli23}).
By using the MCMC method, we obtain fitting results of  $T_{\rm dust} = 50^{+6}_{-5} \rm \ K$, and $\beta = 1.99^{+0.22}_{-0.22}$, $T_{\rm dust} = 116^{+18}_{-11} \rm \ K$, and $\beta = 2.53^{+0.37}_{-0.34}$, and $T_{\rm dust} = 108^{+47}_{-33} \ \rm K$, $\beta = 2.49^{+0.57}_{-0.44}$, and $\lambda_{\rm thick} = 252^{+57}_{-65}\ \mu m$ for Model 1, 2, and 3, respectively (see Fig. \ref{psosed1}). 
The derived dust temperature and emissivity index are in agreement with results obtained for other quasars at $z\sim 6$ adopting an optically thick assumption (e.g., \citealt{wangf19}; \citealt{pensabene22}), as well as results obtained from 3D radiative transfer calculations
\citep{dimascia21}. 
The resulting infrared (IR, 8 $\mu m - 1000\ \mu m$) luminosities ($L_{\rm IR}$) are $1.80^{+0.20}_{-0.20} \times 10^{13}$ \lsun{}, $4.61^{+0.25}_{-0.25} \times 10^{13}$ \lsun{}, and $2.70^{+0.40}_{-0.40} \times 10^{13}$ \lsun{} for the three models, respectively.
If we adopt the previously widely used optically thin modified blackbody dust model for $z\sim 6$ quasars with parameters of $T_{\rm dust} = 47 \rm \ K$, and $\beta = 1.6$ \citep{beelen06}, the derived infrared luminosity is 1.9 $\times 10^{13}$ \lsun{}, which is close to the result of Model 1.

\qsos{} is a radio-loud quasar, we thus also include a power-law radio emission component $f_{\nu} \propto \nu^{\alpha}$, with a power-law index of $\alpha$ to fit the data in addition to the thermal dust SED model (Eq. \ref{eq6}). 
Besides the continuum emission detected in this paper, we include published continuum data at 1.4, 3 GHz from the Very Large Array Sky Survey (VLASS) and Faint Images of the Radio Sky at Twenty Centimeters (FIRST) (\citealt{villarreal18}; \citealt{becker95}), and 32 and 250 GHz obtained using the Karl Jansky Very Large Array (VLA), and NOEMA (\citealt{wang11}; \citealt{khusanova22}), respectively.
The observational data cannot constrain all the model parameters for the radio power-law + general opacity model. We thus fix $\lambda_{\rm thick}=100 \mu m$ (typical of SMGs) and explore two cases with different assumptions of $T_{\rm dust}$: namely, 50 K (Model 1) and 100 K (Model 2) representing a typical ``cold" and ``warm" dust, respectively \footnote{We also considered to fix $\lambda_{\rm thick}=264 \mu m$ (typical of QSOs at $z\sim 6$), but the models could not fit the SED of \qsos{} at 250 GHz.}. 

\begin{figure*}
\gridline{\fig{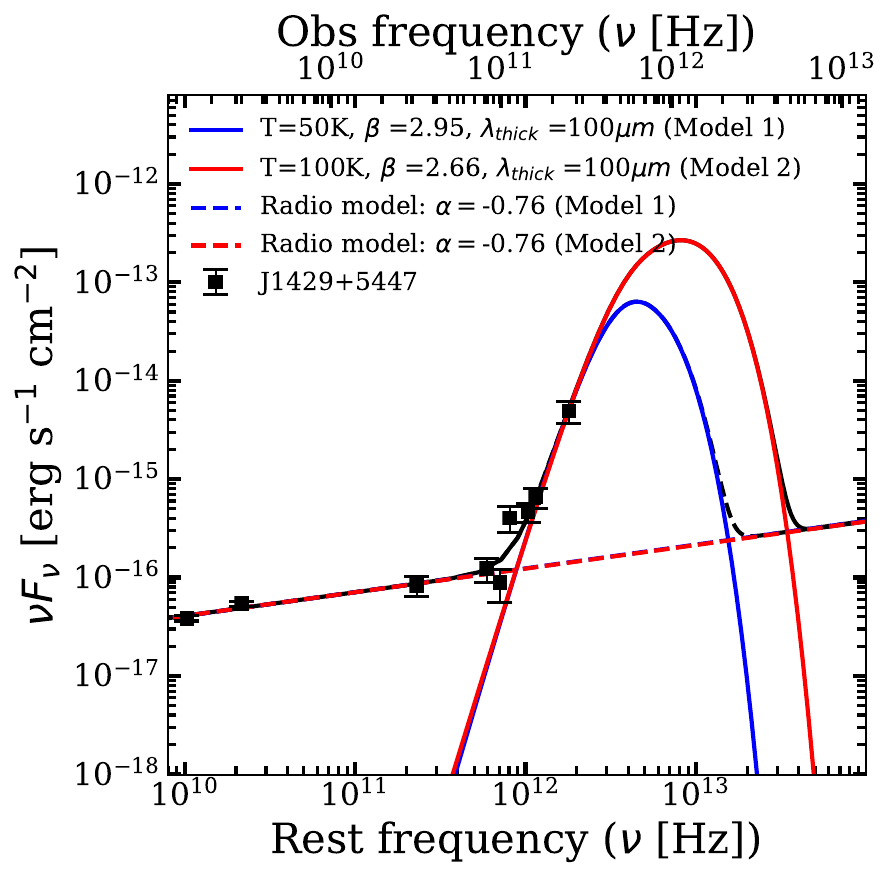}{0.51\textwidth}{(a)}
          }
\gridline{\fig{J1429_dust_bayesian_1.pdf}{0.5\textwidth}{Model 1 }
          \fig{J1429_dust_bayesian_2.pdf}{0.5\textwidth}{Model 2 }
          }
\caption{\label{psosed2} (a) Dust SED fitting results for \qsos{}. The data are shown as black squares. The Top and bottom axes show the observed and rest continuum frequency. The dust SED is fitted with the general opacity dust SED model + radio power-law emission (see Sect. \ref{dustsed} for further details). The solid/dashed blue (red) lines represent the best-fit general opacity/radio power-law models for Model 1(2).
(b) and (c) Posterior distributions of the model parameters for \qso{} (c to e) and \qsos{}, respectively.  The color scheme for different models is the same as in panel (a).   The 2D contours show the [1, 2]$\sigma$ confidence intervals in the marginalized distributions.} 
\end{figure*}

The least square estimation suggests best-fit values of $\alpha = -0.76$, and $\beta = 2.95$ for Model 1, and $\alpha = -0.76$, and $\beta = 2.66$ for Model 2.  The resulting $\tau_{1900GHz} $ are 0.26 and 0.30 for Model 1 and 2, respectively, which are consistent with values derived for other quasars at $z\sim 6$ (\citealt{walter22}; \citealt{decarli23}).
The MCMC method finds $\beta = 3.02^{+0.50}_{-0.53}$ and $\alpha = -0.76^{+0.04}_{-0.05}$, and $\beta = 2.74^{+0.58}_{-0.55}$ and $\alpha = -0.76^{+0.04}_{-0.05}$, for Model 1 and 2, respectively (Fig. \ref{psosed2}). The derived infrared luminosities are $6.89^{+0.18}_{-0.18} \times 10^{12}$ \lsun{} and $3.68^{+0.42}_{-0.42} \times 10^{13}$ \lsun{} for Model 1 and 2, respectively.  If we adopt the widely used modified blackbody dust SED model with parameters of $T_{\rm dust} = 47 \rm \ K$, and $\beta = 1.6$ instead, the infrared luminosity is 9.5 $\times 10^{12}$ \lsun{}.

The derived dust properties of \qso{} and \qsos{} are presented in Table \ref{fit_dust_cont}.
Future high-frequency continuum observations, possibly sampling the peak of the dust SED and/or the Wien tail will be critical to further constrain the dust temperature as well as the IR luminosity.

\begin{deluxetable*}{llll}[ht]
\tablecaption{Dust SED and CO SLED modeling results \label{fit_dust_cont}}
\tablehead{\colhead{Source} & Model  &\colhead{LS Best-fit} & \colhead{MCMC}}
\startdata
Dust SED &Model($\lambda_{\rm thick-fixed}$,$T_{\rm dust-fixed}$) & \\
\qso{} & Model 1($100\mu m$, $\ /$ ) & $T_{\rm dust} = 48 \rm K$, $\beta = 2.05$& $T_{\rm dust} = 50^{+6}_{-5} \rm K$, $\beta = 1.99^{+0.22}_{-0.22}$, $L_{\rm IR}$=$1.80^{+0.20}_{-0.20} \times 10^{13}$ \lsun{}\\
&Model 2($264\mu m$, $\ /$ ) &$T_{\rm dust} = 111 \rm K$, $\beta = 2.62$ &$T_{\rm dust} = 116^{+18}_{-11} \rm K$, $\beta = 2.53^{+0.37}_{-0.34}$, $L_{\rm IR}$=$4.61^{+0.25}_{-0.25} \times 10^{13}$ \lsun{}\\
&Model 3($\ /$ ,$\ /$ )&$T_{\rm dust} = 80 \rm K$, $\beta = 2.33$, $\lambda_{\rm thick} = 199\mu m$ &$T_{\rm dust} = 108^{+47}_{-33} \rm K$, $\beta = 2.49^{+0.57}_{-0.44}$, $\lambda_{\rm thick} = 252^{+57}_{-65}\mu m$, $L_{\rm IR}$=$2.70^{+0.40}_{-0.40} \times 10^{13}$ \lsun{} \\
\qsos{} & Model 1($100 \mu m$, $50 \rm  K$) &  $\alpha = -0.76$, $\beta = 2.95$ &$\beta = 3.02^{+0.50}_{-0.53}$, $\alpha = -0.76^{+0.04}_{-0.05}$, $L_{\rm IR}$=$6.89^{+0.18}_{-0.18} \times 10^{12}$ \lsun{}\\
& Model 2($100 \mu m$, $100 \rm K$) &$\alpha = -0.76$, $\beta = 2.66$ &$\beta = 2.74^{+0.58}_{-0.55}$, $\alpha = -0.76^{+0.04}_{-0.05}$, $L_{\rm IR}$=$3.68^{+0.42}_{-0.42} \times 10^{13}$ \lsun{}\\
\hline 
CO SLED\\
\qso{} & PDR  & \nht{} = $10^{6.0}$,\radpdr{} = $10^{6.0}$ & \nodata  \\
& Extreme PDR& \nht{} = $10^{7.1}$,\radpdr{} =  $10^{7.8}$ & \nodata  \\
&  XDR  &\nht{} = $10^{4.3}$,\radxdr{} = $10^{1.1}$ &  \nht{} = $10^{4.7^{+0.8}_{-0.6}}$, \radxdr{} = $10^{0.87^{+0.6}_{-0.4}}$ \\
\qsos{} & PDR  & \nht{} = $10^{6.0} $, \radpdr{} =  $10^{5.4}$& \nht{} = $10^{5.8^{+0.7}_{-0.3}}$, \radpdr{} = $10^{5.6^{+0.4}_{-0.7}}$  \\
&  XDR  &\nht{} = $10^{3.7}$, \radxdr{} = $10^{1.5}$ &\nht{} = $10^{3.8^{+1.1}_{-0.3}}$, \radxdr{} = $10^{0.33^{+1.4}_{-0.7}}$\\
&  PDR + XDR &PDR: \nht{} = $10^{5.1}$, \radpdr{} = $10^{4.2}$ & \nht{} = $10^{5.1^{+0.7}_{-1.1}}$, \radpdr{} = $10^{4.2^{+0.4}_{-0.5}}$;\\
&   &XDR: \nht{} = $10^{4.3}$, \radxdr{} = $10^{2.0}$ & \nht{} = $10^{4.4^{+0.9}_{-0.7}}$, \radxdr{} = $10^{1.3^{+1.4}_{-1.3}}$\\
\enddata
\tablecomments{The dust SED and the CO SLED fitting results. ``LS Best-fit" represents the best-fit result of the least square method. ``MCMC" represents the fitting result of the MCMC method. Model($\lambda_{\rm thick-fixed}$,$T_{\rm dust-fixed}$) represents the model where $\lambda_{\rm thick}$ or $T_{\rm dust}$ is fixed to $\lambda_{\rm thick-fixed}$ or $T_{\rm dust-fixed}$. 
As for the CO SLED modeling, \nht{},\radpdr{}, and \radxdr{} are in units of $\rm cm^{-3}$, habing field, and $\rm erg\ s^{-1}\ cm^{-2}$. }
\end{deluxetable*}

\subsection{ISM properties}\label{cofit}

CO SLEDs are sensitive probes of the physical conditions (e.g., temperature, density, and radiation field) of the molecular gas. To derive the physical properties of the molecular gas in the host galaxies of these two quasars, we employ the CO SLED grid models published in \cite{pensabene21}. These models consider two different gas heating mechanisms, namely a PDR heated by the FUV photons from young massive stars and an XDR heated by the X-ray photons from the AGNs. 
In the PDR model, the shape of the CO SLED is described by two physical parameters: the gas volume density \nht{} and the FUV radiation field strength \radpdr{} in Habing field units (1 Habing field corresponds to $\rm 1.6 \times 10^{-3}\ erg\ s^{-1}\ cm^{-2}$; \citealt{habing68}). The CO SLED in the XDR model is determined by the gas volume density \nht{} and the strength of the X-ray radiation field  \radxdr{} in the unit of $\rm erg\ s^{-1}\ cm^{-2}$. 
To fully sample the molecular region which has a typical column density of $\textgreater 2 \times 10^{22} \ \rm cm^{-2}$ \citep{mckee07}, we fix the column density of molecular hydrogen to $10^{23} \ \rm cm^{-2}$ for both sets of the PDR and XDR models. The adopted column density is also consistent with the values derived/adopted for other z$\sim$6 quasars (e.g., \citealt{meyer22}; \citealt{decarli23}). 
Ranges of \nht{},  \radpdr{}, and \radxdr{} for typical molecular clouds explored in the model grids are $10^{2}-10^{6} \ \rm cm^{-3}$,   $10^{1}-10^{6}$ and $10^{-2}-10^{2} \rm \ erg\ s^{-1}\ cm^{-2}$ \citep{pensabene21}.  
We consider two cases for the CO SLED fitting: 1) a single PDR component and  2) an XDR component.
The fitting procedure for each case is as follows. Firstly, we use the least square estimation for a first-order estimation of the best-fit values of the parameters. Secondly, we employ the MCMC package emcee to sample the posterior distribution of the parameters while using the best-fit values derived above as educated initial guesses for the parameters \citep{foreman13}.

\begin{figure*}
\gridline{\fig{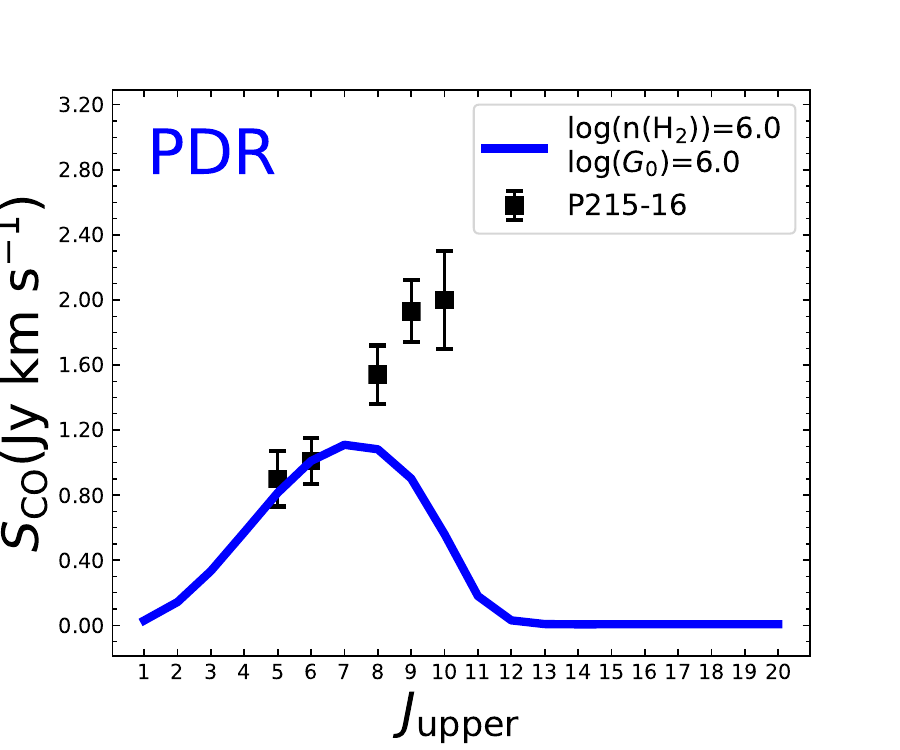}{0.33\textwidth}{(a)}
          \fig{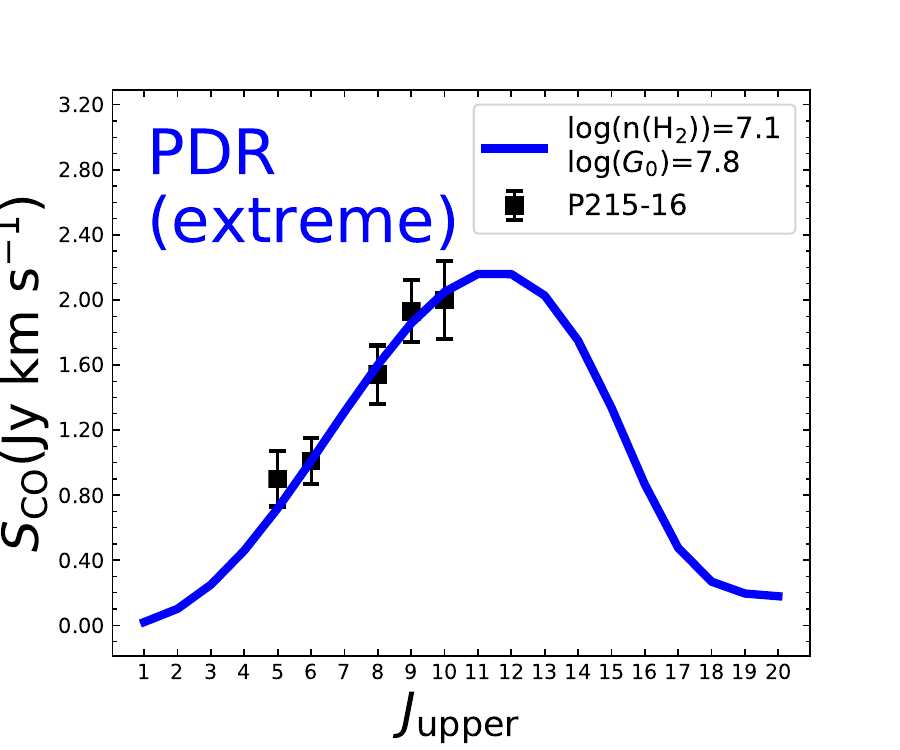}{0.33\textwidth}{(b)}
          \fig{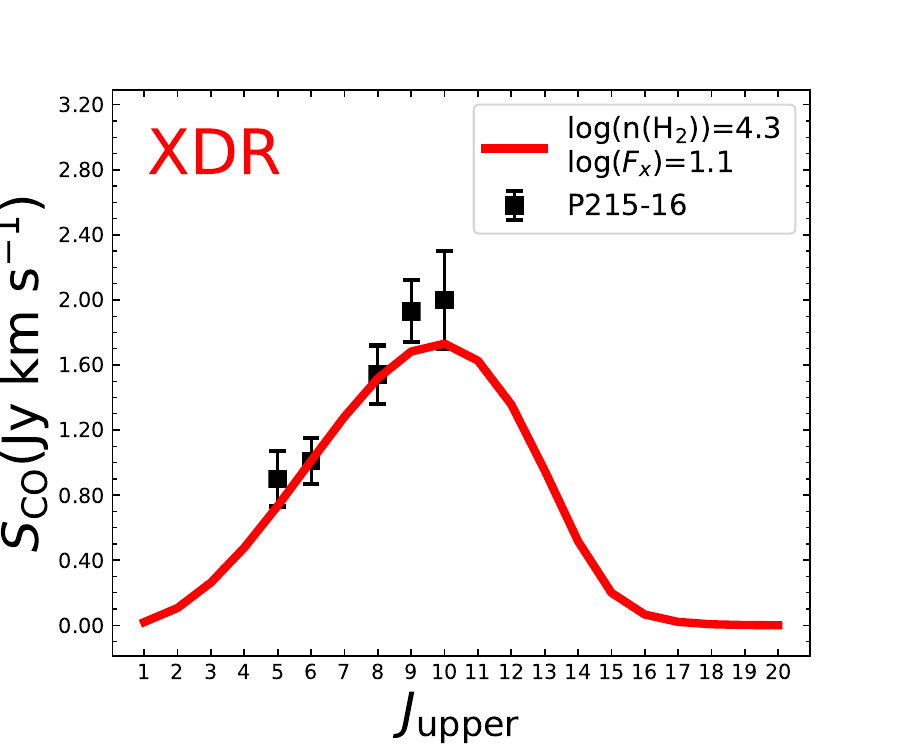}{0.33\textwidth}{(c}
          }
\gridline{\fig{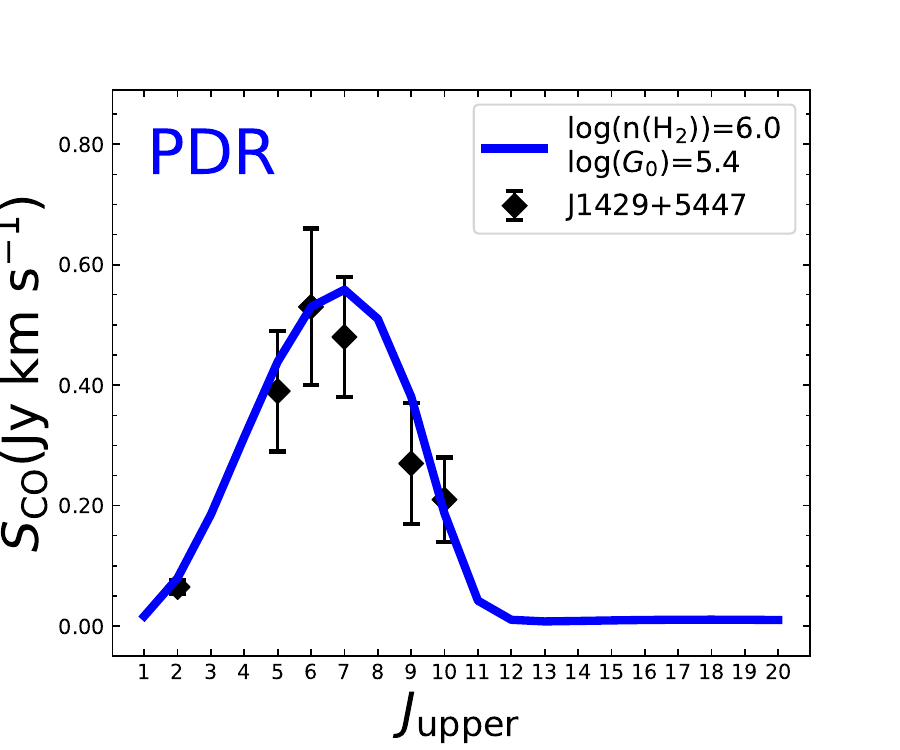}{0.33\textwidth}{(d)}
          \fig{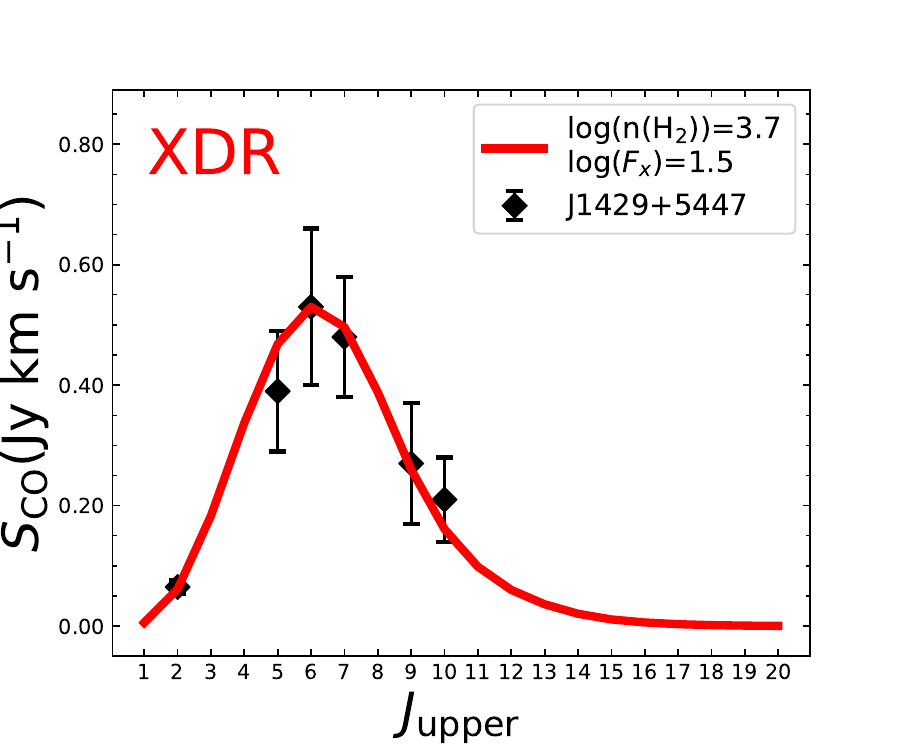}{0.33\textwidth}{(e)}
          \fig{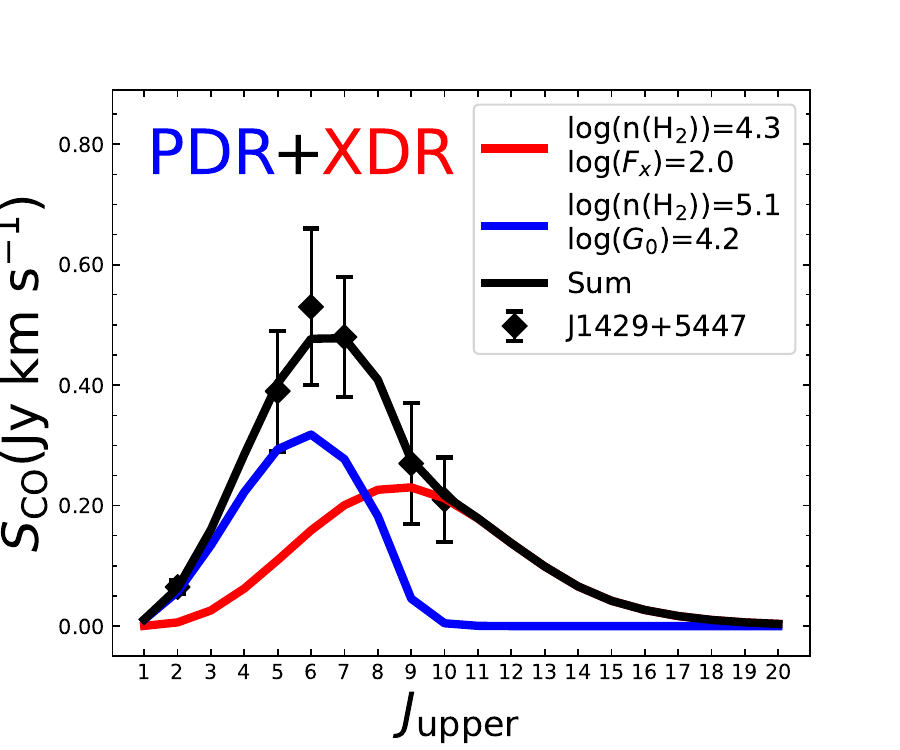}{0.33\textwidth}{(f)}
          }
\caption{\label{cosledpso} CO SLED fitting results of the PDR (left), extreme-PDR (middle), and XDR model (right) for \qso{} (top) and  PDR (left), XDR (middle), and PDR+XDR (right) model for \qsos{} (bottom). The fitting results are shown in solid blue (PDR) and red (XDR) lines, and the data are shown in black squares and diamonds for \qso{} and \qsos{}, respectively. The sum of the PDR+XDR model is shown as a solid black line.} 
\end{figure*}

We show the fitting results in Fig. \ref{cosledpso}. The CO SLED of \qso{} cannot be fitted with any of our PDR models with \nht{} and \radpdr{} in the range of  $10^{2}-10^{6} \ \rm cm^{-3}$ and $10^{1}-10^{6}$ (Fig. \ref{cosledpso}(a)), which are typical ranges of the PDR regions in star-forming galaxies, starbursts and AGNs (e.g., \citealt{gallerani14}; \citealt{pensabene14}; \citealt{pensabene21}; \citealt{esposito22}).
To further explore if the data can be fitted with an extreme PDR model, we extend the range of \nht{} and \radpdr{} parameters of \citet{pensabene21} to $10^{2}-10^{8} \ \rm cm^{-3}$ and $10^{1}-10^{8}$, respectively, while using the same set of assumptions in CLOUDY. 
The results suggest that the data can be fitted with an extreme PDR of  \nht{} = $10^{7.1} \ \rm cm^{-3}$ and \radpdr{} =  $10^{7.8}$ (Fig. \ref{cosledpso}(b)). 
The CO SLED of \qso{} can also be fitted with an XDR model. The least square approximation finds a dense XDR component of \nht{} = $10^{4.3} \ \rm cm^{-3}$ illuminated by a radiation field of \radxdr{} = $10^{1.1}$  $\rm erg\ s^{-1}\ cm^{-2}$. 
The MCMC analysis finds \nht{} = $10^{4.7^{+0.8}_{-0.6}}$ $\rm cm^{-3}$ and \radxdr{} = $10^{0.87^{+0.6}_{-0.4}}$ $\rm erg\ s^{-1}\ cm^{-2}$ for the XDR model, where values and uncertainties are estimated based on the 50th, 16th, and 84th percentiles of the samples in the marginalized distributions of the parameters.
It is also possible that the CO SLED of \qso{} is contributed by both PDR and XDR components. However, the current CO SLED measurements are insufficient to constrain the parameters for a PDR+XDR model. Future observations of the high-$J$ ($J \geq 11$) CO lines will be critical to discriminate between different models. 

\begin{figure}[ht]
\includegraphics[width=\columnwidth]{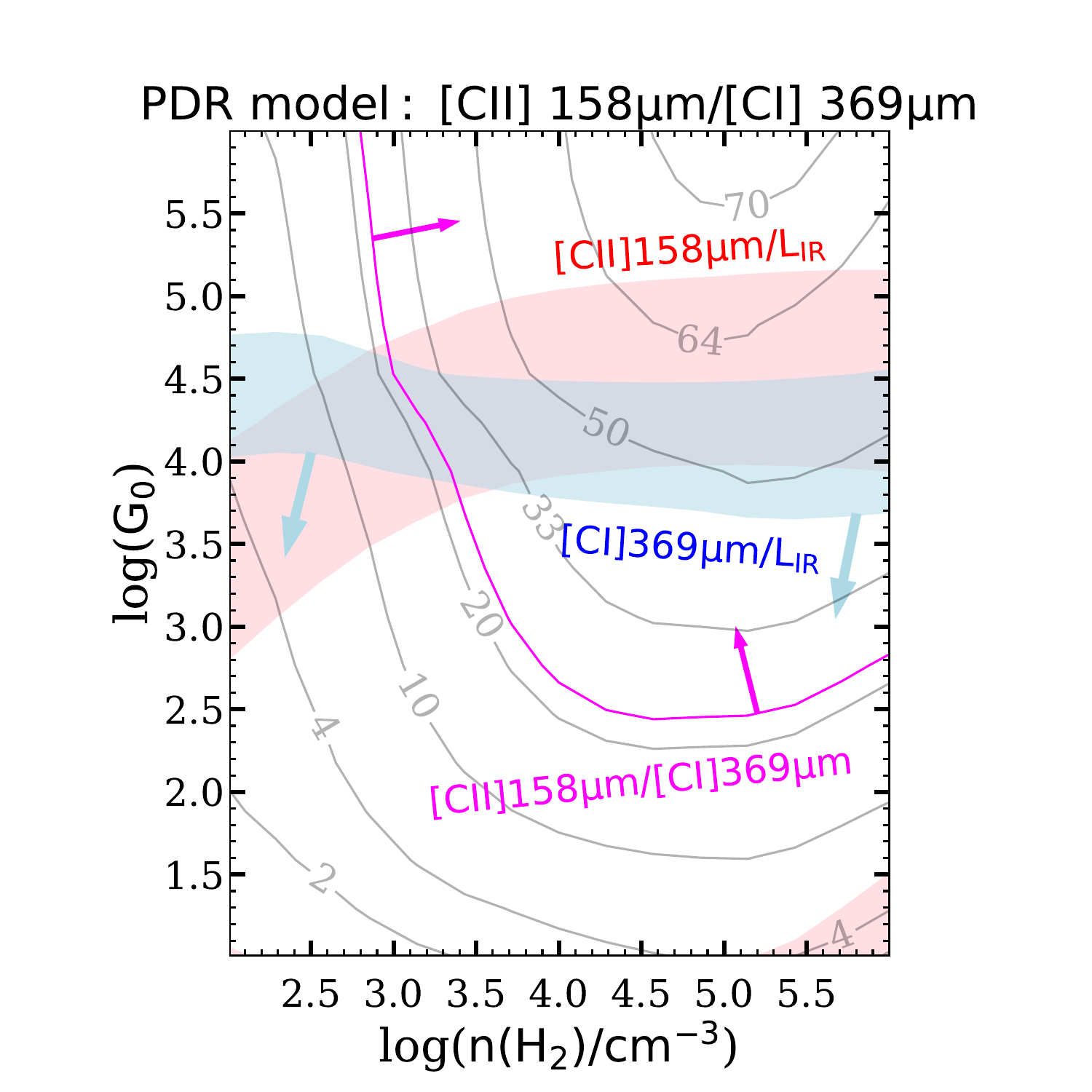} 
  \caption{CLOUDY PDR model prediction for the \cii{}/\ci{} line ratio (grey contours) with different gas density ($\rm n(H_{2})$) and radiation field  ($\rm G_{0}$). We also show the value of the \cii{}/$L_{\rm IR}$ ratio (pink region), the lower-limit of the \cii{}/\ci{} ratio (magenta solid line),  and the upper-limit of the \ci{}/$L_{\rm IR}$ ratio (light blue region) for \qsos{}. }
  \label{ciicigrid}
\end{figure}

As for \qsos{}, the combination of available \cii{} data with the information on the non-detection of the \ci{} line from this work enables us to constrain the physical conditions of the atomic gas. 
The \cii{}/\ci{} luminosity ratio of $> 23$ argues for a PDR origin of the line excitation, as the XDR scenario usually results in a lower line ratio of $\lesssim 10$ \citep{pensabene21}. Predictions of the \cii{}/\ci{} ratio in the PDR model are shown in Fig. \ref{ciicigrid} \citep{pensabene21}. 
The \cii{}/\ci{}, \cii{}/$L_{\rm IR}$, and \ci{}/$L_{\rm IR}$ ratios indicate a PDR component with parameters in the range of \nht{}= $10^{3 - 6}$ $\rm cm^{-3}$ and \radpdr{} = $10^{3.7 - 4.7}$ adopting an $L_{\rm IR}$ value of  $6.89^{+0.18}_{-0.18} \times 10^{12}$ \lsun{} $-$ $3.68^{+0.42}_{-0.42} \times 10^{13}$ \lsun{}.
Detections of the CO emission lines complemented with previous \cotwo{} observations \citep{wang11} enable us to constrain, for the first time, the molecular ISM physical conditions of a radio-loud quasar at $z> 6$. 

We analyze the CO SLED of \qsos{} following a similar procedure as that of \qso{}. 
A single ``dense'' PDR model illuminated by an intense FUV radiation field is able to reproduce the observational data, with parameters of \nht{} = $10^{6.0} \ \rm cm^{-3}$, \radpdr{} =  $10^{5.4}$. 
A single XDR model is also able to reproduce the CO SLED with best-fit parameters of \nht{} = $10^{3.7} \ \rm cm^{-3}$, \radxdr{} = $10^{1.5}$  $\rm erg\ s^{-1}\ cm^{-2}$. 
Using the gas conditions derived from \cii{}, \ci{} and $L_{\rm IR}$ as a constraint of the PDR component, we also try to fit the CO SLED of \qsos{} with a PDR + XDR model. 
This leads to a best-fit result of \nht{} = $10^{4.3}$ $\rm cm^{-3}$ and \radxdr{} = $10^{2.0}$ $\rm erg\ s^{-1}\ cm^{-2}$ for the XDR component, and  \nht{} = $10^{5.1}$ $\rm cm^{-3}$ and \radpdr{} = $10^{4.2}$ for the PDR component. The PDR component contributes to 96$\%$ of the total molecular gas mass in this case. 
The MCMC method finds best-fit values in the range of \nht{} = $10^{5.8^{+0.7}_{-0.3}}$ $\rm cm^{-3}$ and \radpdr{} = $10^{5.6^{+0.4}_{-0.7}}$ (PDR model), \nht{} = $10^{3.8^{+1.1}_{-0.3}}$ $\rm cm^{-3}$ and \radxdr{} = $10^{0.33^{+1.4}_{-0.7}}$ $\rm erg\ s^{-1}\ cm^{-2}$ (XDR model), and \nht{} = $10^{5.1^{+0.7}_{-1.1}}$ $\rm cm^{-3}$ and \radpdr{} = $10^{4.2^{+0.4}_{-0.5}}$, and \nht{} = $10^{4.4^{+0.9}_{-0.7}}$ $\rm cm^{-3}$ and \radxdr{} = $10^{1.3^{+1.4}_{-1.3}}$ $\rm erg\ s^{-1}\ cm^{-2}$   (PDR + XDR model).
Analysis of the gas properties using either the \cii{}, \ci{} and IR ratios or the CO SLED suggests that the \cii{}, \ci{}, CO and IR emission can be explained with one PDR component. However, we are not able to rule out a single XDR model or a composite PDR + XDR model, which reproduces the observed CO SLED. In the case of either XDR or PDR + XDR model, a ``moderate density" XDR component illuminated by an intense radiation field is suggested. 

Results of the gas properties derived from CO SLED modeling are shown in Table \ref{fit_dust_cont}.
In short, both the extreme-PDR and the XDR model could explain the observed CO SLEDs (up to $J$ = 10) of \qso{}. However, we also notice that the extreme-PDR model for \qso{} requires extremely high gas density and FUV radiation field intensity which is rarely seen in starburst systems and may suggest AGN contribution. For \qsos{}, the CO SLED could be fitted either with a PDR, XDR, or PDR + XDR model. 
More discussions about the excitation mechanism of the molecular gas can be found in Section \ref{discussion}.

In addition, we robustly detect the \htot{} line in the quasar \qso{}, and tentatively detect the \htoo{} in \qso{} and \htoo{} and \htot{} in \qsos{}. The \hto{} molecule, as the third most abundant species after $\rm H_{2}$ and CO in the molecular ISM phase, traces obscured star formation in dense molecular clouds. Linear relations were reported between the mid/high-$J$ \hto{} and the IR luminosities in local and high-$z$ infrared bright galaxies and AGNs (e.g., \citealt{yang13,yang16}; \citealt{pensabene22}). The detections and upper limits of \htoo{} and \htot{} for \qso{} and \qsos{} follow the \hto{}$-$IR trend of other $z\sim 6$ quasars (\citealt{yang19}; \citealt{pensabene22}; Riechers et al. in prep).

\section{Discussion: The heating mechanisms of the molecular gas \label{discussion} }
With the NOEMA observations of the multiple CO transitions, we analyze the CO SLEDs of two $z > 5.7$ quasars. Both of them are amongst the most infrared luminous objects found so far in the early Universe.  The IR luminosities are of order of a few $10^{12}$ to $10^{13}$ \lsun{}, revealing massive star formation in a dusty ISM.  \qsos{} is a radio-loud quasar, which enables us to study the CO excitation close to a young and radio-loud AGN for the first time.

\begin{figure*}
\includegraphics[width=1.0\textwidth]{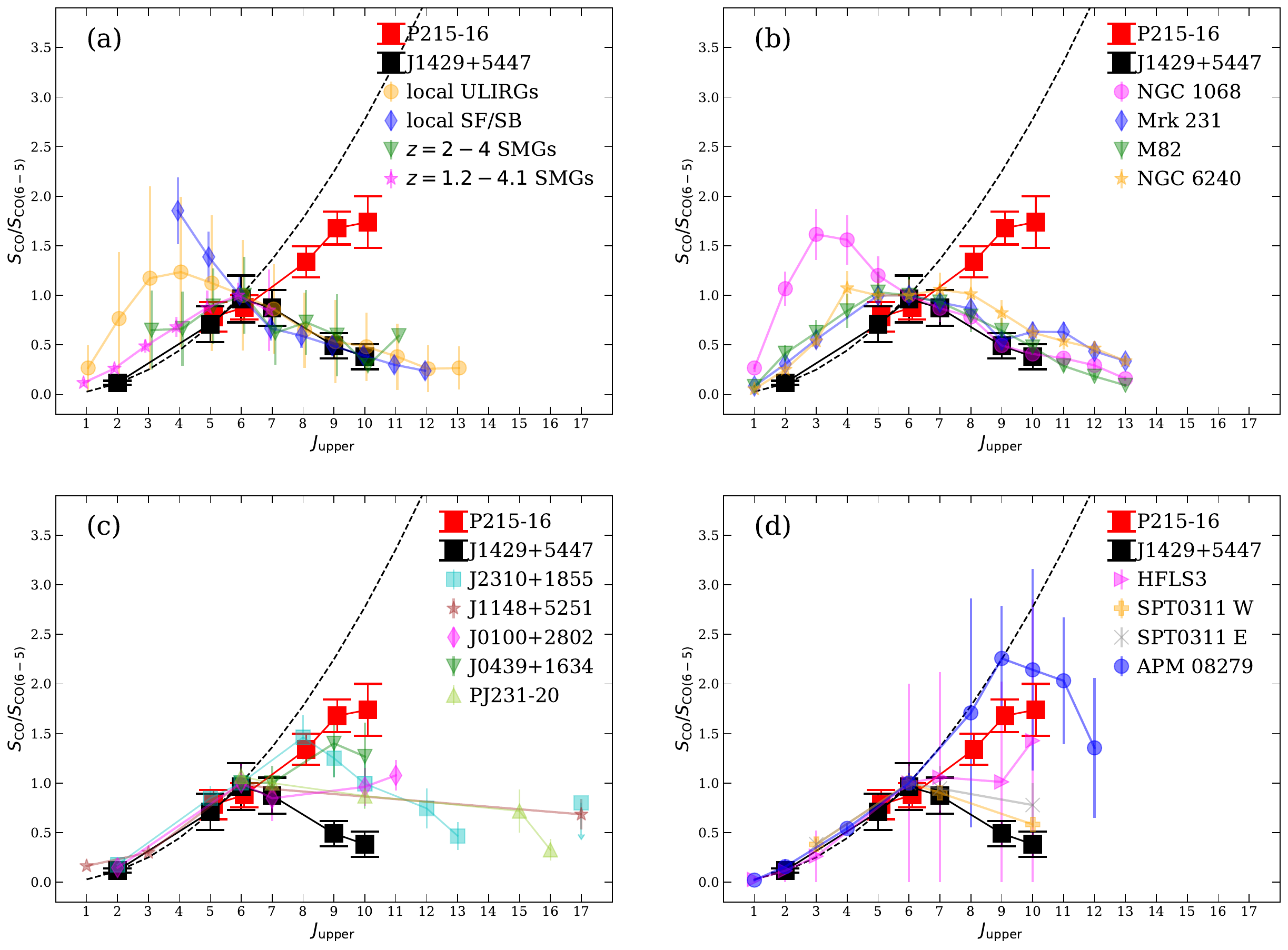}
\caption{\label{cosledcompare} CO SLEDs of \qso{} and \qsos{} in comparison with (a) the mean value of local and high-$z$ starburst galaxy samples, (b) local representative AGNs and starburst galaxies, (c) other quasars at $z\gtrsim 6$,  and (d) starburst galaxies at $z\gtrsim 6$ and the lensed quasar APM 08279 at $z=3.9$. The dashed line represents the thermalized CO excitation with constant brightness temperature on the Rayleigh–Jeans scale, i.e., $L^{'}_{CO(J-J-1)}$ = $L^{'}_{CO(1-0)}$.}
\end{figure*}

To further explore the possible origins of the different CO excitations observed in \qso{} and \qsos{}, we compare our observations with the CO SLEDs of local and high-$z$ starburst galaxies and AGNs. The results are shown in Fig. \ref{cosledcompare}. 
The local and high-$z$ galaxy samples are local (ultra-) luminous infrared galaxies (``local (U)LIRGs"; \citealt{rosenberg15}), local star-forming and starburst galaxies (``local SF + SB"; \citealt{liu15}), lensed (sub)millimeter galaxies at $z=2-4$ (``$z=2-4$ SMGs"; \citealt{yang17}), and (sub)millimeter galaxies at $z=1.2-4.1$ (`` $z=1.2-4.1$ SMGs"; \citealt{bothwell13}; Fig. \ref{cosledcompare}a).  
 The representative local galaxies comprise three AGNs, NGC 1068 \citep{spinoglio12}, Mrk 231 \citep{van10}, and NGC 6240 \citep{rosenberg15}, and the starburst galaxy M82 (\citealt{weiss05}; \citealt{panuzzo10}; Fig. \ref{cosledcompare}b). 
The $z\gtrsim 6$ comparison sources include the $z\sim 6$ quasars that have previous published CO SLEDs with detections of at least four transitions: J2310+1855 \citep{li20}, J1148+5251 (\citealt{bertoldi03}; \citealt{walter03}; \citealt{beelen06}; \citealt{riechers09}; \citealt{gallerani14}), J0100+2802 \citep{wangf19}, J0439+1634 \citep{yang19}, and PJ231--20 (\citealt{pensabene21}; Fig. \ref{cosledcompare}c). As the \cosix{} line is not observed in PJ231--20, we normalize its CO SLED using the predicted \cosix{} flux from the best fit radiative transfer model of \cite{pensabene21}. 
In addition, we collect starburst galaxies at $z \gtrsim 6$, including SPT 0311 W, SPT 0311 E \citep{jarugula21}, and HFLS3 \citep{riechers13}, as well a lensed quasar APM 08279 at $z=3.9$ which reveals the highest excited CO SLED ever published across redshifts (\citealt{weiss07}; \citealt{braford11}; Fig. \ref{cosledcompare}d). 

The CO SLEDs of local and high-$z$ starburst galaxy samples typically peak at $J \lesssim 6$, and drop rapidly at high $J$, resulting in low line fluxes at $J > 10$ (Fig. \ref{cosledcompare}a). This is consistent with the gas heated by the FUV photons from young massive stars, indicating a PDR origin.
On the other hand, the local AGNs and high-$z$ quasars tend to have higher CO excitation with the CO SLEDs peaking at $J \gtrsim 6$. The bright high-$J$ ($J\gtrsim 9$) CO transitions are usually associated with other heating mechanisms, e.g., X-ray heating from AGNs, cosmic ray heating, and mechanical heating by shocks in addition to the FUV heating from young massive stars (Fig. \ref{cosledcompare}b and c; e.g., \citealt{van10}; \citealt{spinoglio12}). 
As is suggested by Fig. \ref{cosledcompare} and radiative transfer model predictions, the CO SLEDs of the starburst galaxies and AGNs/quasars are distinguishable through observations of the high-$J$ (e.g., $J\gtrsim 9$) CO transitions (e.g., \citealt{vallini19}). 

\qso{} exhibits a highly excited CO SLED indicating an excitation level that is much higher than that of all local/high-$z$ starburst galaxies (Fig. \ref{cosledcompare}a and b) and local AGNs (Fig. \ref{cosledcompare}b), with no turn-over up to $J=10$.  It has the highest \coten{}/\cosix{} ratio among all published quasars at $z\sim 6$ (Fig. \ref{cosledcompare}c) and the second highest excited CO ever across redshifts (less excited compared to APM 08279).
We have demonstrated in Section \ref{cofit} that either an XDR component with high gas density and intense radiation (\nht{} = $10^{4.3} \ \rm cm^{-3}$, \radxdr{} = $10^{1.1}$  $\rm erg\ s^{-1}\ cm^{-2}$) or an extreme PDR of  \nht{} = $10^{7.1} \ \rm cm^{-3}$ and \radpdr{} =  $10^{7.8}$ could explain such a highly excited CO SLED (Fig. \ref{cosledpso}). 
Mechanical heating by shocks can also result in highly excited molecular CO which was reported in some local galaxies \citep{rosenberg15}. Shocks are relatively more efficient in heating the gas compared to the dust, and a typical signature of shock-heated gas is a high CO/IR luminosity ratio of a few times $10^{-4}$ (e.g., \citealt{pellegrini13}).  The CO/IR ratio (even for the brightest \coten{} transition) of \qso{} is $\sim 10^{-5}$, which disfavors the shock heating scenario. Enhanced cosmic ray heating could also be responsible for high-$J$ CO line excitation which is sometimes indistinguishable from XDR. 
\cite{vallini19} investigated several CO excitation mechanisms in high-$z$ galaxies and reported that the most extreme cosmic-ray heated CO SLED peaked at $J=9$. The high CO excitation of \qso{}, which peaks at $J\geq 10$, is best explained by an XDR or extreme-PDR component. In either case, AGN could be the source powering the high CO excitation, by contributing to the X-ray emission or intense FUV emission as suggested by the results of the XDR and extreme-PDR model.

To the contrary, \qsos{} exhibits a relatively lower CO excitation, which is comparable to that measured for local and high-$z$ starburst galaxies (Fig. \ref{cosledcompare}a and b), and $z \gtrsim 6$ starburst galaxies, e.g., SPT 0311 W, SPT 0311 E, and HFLS3 (Fig. \ref{cosledcompare}d). 
It has the lowest CO$ (J - J$-1)$_{[J \geq 7]}$ /\cosix{} line ratios among the $z \sim 6$ quasars with published CO SLEDs (Fig. \ref{cosledcompare}d). 
The CO SLED could be explained by a single PDR model with extreme gas densities of \nht{} = $10^{6.0} \ \rm cm^{-3}$ and radiation of \radpdr{} =  $10^{5.4}$.

We also explore if the CO SLED of \qsos{} can be fitted with an XDR model included, given the luminous X-ray detection in this quasar. 
The CO SLED of \qsos{} fitted by an XDR or a PDR+XDR model prefers an XDR component with moderate density (\nht{} = $10^{3.7-4.3} \ \rm cm^{-3}$) and illuminated by an intense X-ray radiation field (\radxdr{} = $10^{1.5-2.0}$  $\rm erg\ s^{-1}\ cm^{-2}$).  
This is at the lower edge of the molecular gas density range found in $z\sim 6$ quasars (e.g., \citealt{li20}; \citealt{pensabene21}). As \qsos{} is a radio-loud object, it is possible that the radio jet expels the gas out of the host galaxy resulting in lower gas density. This is consistent with the indication of a possible outflow found in previous \cii{} observations \citep{khusanova22}. The outflowing gas mass constrained from the broad \cii{} component is $1.2 \times 10^{10}$ \msun{} assuming an $\alpha_{\rm [CII]} = 30 \rm \  M_{\sun}\ L_{\sun}^{-1}$ \citep{zanella18}.
The \oh{} absorption is a sensitive tracer of outflows in galaxies and quasars (e.g.,  \citealt{shao22}). Unfortunately, the low S/N of our \oh{} spectrum prevents us from deriving useful constraints on the outflowing molecular gas mass and deeper observations will be needed to obtain an independent estimate.
In this work, we only include a simple plane parallel geometry and include no dust torus attenuation. A low CO excitation can be produced in a sophisticated geometry XDR model of high dust attenuation even when the X-ray flux and gas density are high (e.g., \citealt{vallini19}). 
Future CO SLED observations of more radio-loud quasars will possibly enable us to study if the low excitation is rare/common in the radio-loud population.


\section{Summary}
We present NOEMA observations and report detections of multiple CO transitions (from \cofive{} to \coten{}), \htot{} and the underlying continuum emission from the host galaxies of the quasars \qso{} at \reds{} and \qsos{} at \redss{}.

Adopting a general opacity dust SED model, we derive dust temperatures in the range of $\sim 50-110\ \rm K$ for the two quasar host galaxies. The derived infrared luminosities of \qso{} and \qsos{} differ by factors of $\sim$ 2.5 and 5.3, respectively, depending on the best-fit models used.

The quasar \qso{} exhibits highly excited molecular CO with a CO SLED peaking at $ J \geq 10$. It has the highest CO$ (J - J$-1)$_{[J \geq 7]}$ /\cosix{} ratio among the $z\sim 6$ quasars with published CO SLEDs known to date. 
Analysis of the CO SLED with models suggests that a ``dense" (\nht{} = $10^{4.3} \ \rm cm^{-3}$) XDR component illuminated by an intense X-ray radiation field (\radxdr{} = $10^{1.1}$) or a ```dense" PDR component with \nht{} = $10^{7.1} \ \rm cm^{-3}$ and \radpdr{} =  $10^{7.8}$  reproduce the observed CO SLED. In either case, the central AGN is likely to contribute to the radiation field to excite the high$-J$ CO transitions.

The CO SLED of the radio-loud quasar \qsos{} reveals the lowest CO excitation among all published quasar CO SLEDs at $z\sim 6$. 
This is the first reported CO SLED of a radio-loud quasar at this redshift and is comparable to those of local AGNs and starburst galaxies at $z\gtrsim 6$. 
The CO SLED could be explained by a PDR component.
In addition, we also propose two possibilities where XDR is considered to explain the low CO excitation: 1) a luminous X-ray radiation field illuminating a moderate-density gas, which may be attributed to a result of the gas expelled by an outflow, and 2) a sophisticated geometry XDR model taking into account the X-ray attenuation by a dust torus. In this case, low CO excitation can be reproduced in a high dust attenuation model even when the X-ray fluxes and gas densities are high. 


In addition, we detect bright \htot{} emission in \qso{}. This places \qso{} exactly on the linear relation between the \htot{} and IR luminosities recently found for other quasars at $z\sim 6$ and local/high-$z$ infrared bright galaxies.

Future observations of the CO SLEDs towards more radio-loud quasars at $z\gtrsim 6$, possibly up to ($J\gtrsim$ 10) will be critical to systematically reveal the physical conditions and heating mechanisms of the molecular gas in the host galaxies of the radio-loud quasar population in the early universe. Comparisons between the radio-loud and radio-quiet populations will shed light on the possible different/similar impact of the radio-quiet (loud) AGN on the molecular gas properties.

\section{acknowledgment} 
\begin{acknowledgments}
This work is based on observations carried out under project numbers W0C3, W18EE and, subsequently S20CW with the IRAM NOEMA Interferometer. 
IRAM is supported by INSU/CNRS (France), MPG (Germany) and IGN (Spain). 
We gratefully acknowledge the support from the Shuimu Tsinghua Scholar Program of Tsinghua University.
AP acknowledges support from Fondazione Cariplo grant no. 2020-0902.
Yali Shao acknowledges supports from the National Natural Science Foundation of China (NSFC) with Grant No. 12303001 and the Beihang University grant with No. ZG216S2305.
This work has been supported by the National Key R$\&$D Program of China (grant no. 2018YFA0404503), the National Science Foundation of China (grant no. 12073014), and the science research grants from the China Manned Space Project with No. CMS-CSST2021-A05.

\end{acknowledgments}


\begin{thebibliography}{}
\bibitem[Beelen et al.(2006)]{beelen06} Beelen, A., Cox, P., Benford, D.~J., et al.\ 2006, \apj, 642, 694
\bibitem[Becker et al.(1995)]{becker95} Becker, R.~H., White, R.~L., \& Helfand, D.~J.\ 1995, \apj, 450, 559. doi:10.1086/176166
\bibitem[Bertoldi et al.(2003)]{bertoldi03} Bertoldi, F., Carilli, C.~L., Cox, P., et al.\ 2003, \aap, 406, L55

\bibitem[Bothwell et al.(2013)]{bothwell13} Bothwell, M.~S., Smail, I., Chapman, S.~C., et al.\ 2013, \mnras, 429, 3047 
\bibitem[Bradford et al.(2011)]{braford11} Bradford, C.~M., Bolatto, A.~D., Maloney, P.~R., et al.\ 2011, \apjl, 741, L37. doi:10.1088/2041-8205/741/2/L37
\bibitem[Carilli \& Walter(2013)]{carilli13} Carilli, C.~L. \& Walter, F.\ 2013, \araa, 51, 105. doi:10.1146/annurev-astro-082812-140953

\bibitem[Circosta et al.(2021)]{circosta21} Circosta, C., Mainieri, V., Lamperti, I., et al.\ 2021, \aap, 646, A96. doi:10.1051/0004-6361/202039270
\bibitem[da Cunha et al.(2021)]{dacunha21} da Cunha, E., Hodge, J.~A., Casey, C.~M., et al.\ 2021, \apj, 919, 30. doi:10.3847/1538-4357/ac0ae0
\bibitem[De Rosa et al.(2014)]{derosa14} De Rosa, G., Venemans, B.~P., Decarli, R., et al.\ 2014, \apj, 790, 145. doi:10.1088/0004-637X/790/2/145
\bibitem[Decarli et al.(2017)]{decarli17} Decarli, R., Walter, F., Venemans, B.~P., et al.\ 2017, \nat, 545, 457. doi:10.1038/nature22358
\bibitem[Decarli et al.(2018)]{decarli18} Decarli, R., Walter, F., Venemans, B.~P., et al.\ 2018, \apj, 854, 97. doi:10.3847/1538-4357/aaa5aa
\bibitem[Decarli et al.(2023)]{decarli23} Decarli, R., Pensabene, A., Diaz-Santos, T., et al.\ 2023, \aap, 673, A157. doi:10.1051/0004-6361/202245674
\bibitem[Di Mascia et al.(2021)]{dimascia21} Di Mascia, F., Gallerani, S., Behrens, C., et al.\ 2021, \mnras, 503, 2349. doi:10.1093/mnras/stab528

\bibitem[Esposito et al.(2022)]{esposito22} Esposito, F., Vallini, L., Pozzi, F., et al.\ 2022, \mnras, 512, 686. doi:10.1093/mnras/stac313
\bibitem[Foreman-Mackey et al.(2013)]{foreman13} Foreman-Mackey, D., Hogg, D.~W., Lang, D., et al.\ 2013, \pasp, 125, 306. doi:10.1086/670067
\bibitem[Frey et al.(2011)]{frey11} Frey, S., Paragi, Z., Gurvits, L.~I., et al.\ 2011, arXiv:1105.2371
\bibitem[Gallerani et al.(2014)]{gallerani14} Gallerani, S., Ferrara, A., Neri, R., \& Maiolino, R.\ 2014, \mnras, 445, 2848 
\bibitem[Carniani et al.(2019)]{carniani19} Carniani, S., Gallerani, S., Vallini, L., et al.\ 2019, \mnras, 489, 3939. doi:10.1093/mnras/stz2410
\bibitem[Gilli et al.(2022)]{gilli22} Gilli, R., Norman, C., Calura, F., et al.\ 2022, arXiv:2206.03508
\bibitem[Gonz{\'a}lez-Alfonso et al.(2014)]{gonz14} Gonz{\'a}lez-Alfonso, E., Fischer, J., Aalto, S., et al.\ 2014, \aap, 567, A91. doi:10.1051/0004-6361/201423980
\bibitem[Greve et al.(2014)]{greve14} Greve, T.~R., Leonidaki, I., Xilouris, E.~M., et al.\ 2014, \apj, 794, 142. doi:10.1088/0004-637X/794/2/142
\bibitem[Guilloteau \& Lucas(2000)]{guilloteau2000} Guilloteau, S. \& Lucas, R.\ 2000, Imaging at Radio through Submillimeter Wavelengths, 217, 299
\bibitem[Habing(1968)]{habing68} Habing, H.~J.\ 1968, \bain, 19, 421

\bibitem[Jarugula et al.(2021)]{jarugula21} Jarugula, S., Vieira, J.~D., Weiss, A., et al.\ 2021, \apj, 921, 97. doi:10.3847/1538-4357/ac21db
\bibitem[Kamenetzky et al.(2016)]{kamenetzky16} Kamenetzky, J., Rangwala, N., Glenn, J., et al.\ 2016, \apj, 829, 93. doi:10.3847/0004-637X/829/2/93
\bibitem[Kawamuro et al.(2020)]{kawamuro20} Kawamuro, T., Izumi, T., Onishi, K., et al.\ 2020, \apj, 895, 135. doi:10.3847/1538-4357/ab8b62

\bibitem[Khusanova et al.(2022)]{khusanova22} Khusanova, Y., Ba{\~n}ados, E., Mazzucchelli, C., et al.\ 2022, \aap, 664, A39. doi:10.1051/0004-6361/202243660
\bibitem[Khorunzhev et al.(2021)]{khorunzhev21} Khorunzhev, G.~A., Meshcheryakov, A.~V., Medvedev, P.~S., et al.\ 2021, Astronomy Letters, 47, 123. doi:10.1134/S1063773721030026

\bibitem[Kormendy \& Ho(2013)]{kormendy13} Kormendy, J. \& Ho, L.~C.\ 2013, \araa, 51, 511. doi:10.1146/annurev-astro-082708-101811

\bibitem[Li et al.(2020)]{liq20} Li, Q., Wang, R., Fan, X., et al.\ 2020, \apj, 900, 12. doi:10.3847/1538-4357/aba52d
\bibitem[Li et al.(2020a)]{li20} Li, J., Wang, R., Riechers, D., et al.\ 2020, \apj, 889, 162. doi:10.3847/1538-4357/ab65fa
\bibitem[Li et al.(2020b)]{li20b} Li, J., Wang, R., Cox, P., et al.\ 2020, \apj, 900, 131. doi:10.3847/1538-4357/ababac
\bibitem[Liu et al.(2015)]{liu15} Liu, D., Gao, Y., Isaak, K., et al.\ 2015, \apjl, 810, L14
\bibitem[Lu et al.(2014)]{lu14} Lu, N., Zhao, Y., Xu, C.~K., et al.\ 2014, \apjl, 787, L23. doi:10.1088/2041-8205/787/2/L23
\bibitem[McKee \& Ostriker(2007)]{mckee07} McKee, C.~F. \& Ostriker, E.~C.\ 2007, \araa, 45, 565. doi:10.1146/annurev.astro.45.051806.110602
\bibitem[Medvedev et al.(2020)]{medvedev20} Medvedev, P., Sazonov, S., Gilfanov, M., et al.\ 2020, \mnras, 497, 1842. doi:10.1093/mnras/staa2051
\bibitem[Meijerink \& Spaans(2005)]{meijerink05} Meijerink, R. \& Spaans, M.\ 2005, \aap, 436, 397. doi:10.1051/0004-6361:20042398
\bibitem[Meijerink et al.(2007)]{meijerink07} Meijerink, R., Spaans, M., \& Israel, F.~P.\ 2007, \aap, 461, 793. doi:10.1051/0004-6361:20066130
\bibitem[Meijerink et al.(2013)]{meijerink13} Meijerink, R., Kristensen, L.~E., Wei{\ss}, A., et al.\ 2013, \apjl, 762, L16. doi:10.1088/2041-8205/762/2/L16
\bibitem[Meyer et al.(2022)]{meyer22} Meyer, R.~A., Walter, F., Cicone, C., et al.\ 2022, \apj, 927, 152. doi:10.3847/1538-4357/ac4e94
\bibitem[Migliori et al.(2023)]{migliori23} Migliori, G., Siemiginowska, A., Sobolewska, M., et al.\ 2023, \mnras, 524, 1087. doi:10.1093/mnras/stad1959
\bibitem[Mingozzi et al.(2018)]{mingozzi18} Mingozzi, M., Vallini, L., Pozzi, F., et al.\ 2018, \mnras, 474, 3640. doi:10.1093/mnras/stx3011
 \bibitem[Montoya Arroyave et al.(2023)]{montoya23} Montoya Arroyave, I., Cicone, C., Makroleivaditi, E., et al.\ 2023, \aap, 673, A13. doi:10.1051/0004-6361/202245046
\bibitem[Morganson et al.(2012)]{morganson12} Morganson, E., De Rosa, G., Decarli, R., et al.\ 2012, \aj, 143, 142. doi:10.1088/0004-6256/143/6/142


\bibitem[Panuzzo et al.(2010)]{panuzzo10} Panuzzo, P., Rangwala, N., Rykala, A., et al.\ 2010, \aap, 518, L37 
\bibitem[Pereira-Santaella et al.(2014)]{pensabene14} Pereira-Santaella, M., Spinoglio, L., van der Werf, P.~P., et al.\ 2014, \aap, 566, A49. doi:10.1051/0004-6361/201423430
\bibitem[Pensabene et al.(2020)]{pensabene20} Pensabene, A., Carniani, S., Perna, M., et al.\ 2020, \aap, 637, A84. doi:10.1051/0004-6361/201936634
\bibitem[Pensabene et al.(2021)]{pensabene21} Pensabene, A., Decarli, R., Ba{\~n}ados, E., et al.\ 2021, \aap, 652, A66. doi:10.1051/0004-6361/202039696
\bibitem[Pensabene et al.(2022)]{pensabene22} Pensabene, A., van der Werf, P., Decarli, R., et al.\ 2022, \aap, 667, A9. doi:10.1051/0004-6361/202243406
\bibitem[Pellegrini et al.(2013)]{pellegrini13} Pellegrini, E.~W., Smith, J.~D., Wolfire, M.~G., et al.\ 2013, \apjl, 779, L19. doi:10.1088/2041-8205/779/2/L19
\bibitem[Pozzi et al.(2017)]{pozzi17} Pozzi, F., Vallini, L., Vignali, C., et al.\ 2017, \mnras, 470, L64. doi:10.1093/mnrasl/slx077

\bibitem[Riechers et al.(2006)]{riechers06} Riechers, D.~A., Walter, F., Carilli, C.~L., et al.\ 2006, \apj, 650, 604. doi:10.1086/507014

\bibitem[Riechers et al.(2009)]{riechers09} Riechers, D.~A., Walter, F., Carilli, C.~L., \& Lewis, G.~F.\ 2009,\apj, 690, 463 
\bibitem[Riechers et al.(2011a)]{riechers11a} Riechers, D.~A., Carilli, C.~L., Maddalena, R.~J., et al.\ 2011, \apjl, 739, L32. doi:10.1088/2041-8205/739/1/L32
\bibitem[Riechers(2011b)]{riechers11b} Riechers, D.~A.\ 2011, \apj, 730, 108. doi:10.1088/0004-637X/730/2/108
\bibitem[Riechers et al.(2013)]{riechers13} Riechers, D.~A., Bradford, C.~M., Clements, D.~L., et al.\ 2013, \nat, 496, 329. doi:10.1038/nature12050
\bibitem[Ramos Almeida et al.(2022)]{ramos22} Ramos Almeida, C., Bischetti, M., Garc{\'\i}a-Burillo, S., et al.\ 2022, \aap, 658, A155. doi:10.1051/0004-6361/202141906

\bibitem[Rosenberg et al.(2015)]{rosenberg15} Rosenberg, M.~J.~F., van der Werf, P.~P., Aalto, S., et al.\ 2015, apj, 801, 72 

\bibitem[Ramos Almeida et al.(2022)]{ramos22} Ramos Almeida, C., Bischetti, M., Garc{\'\i}a-Burillo, S., et al.\ 2022, \aap, 658, A155. doi:10.1051/0004-6361/202141906
\bibitem[Riechers et al.(2013)]{riechers13} Riechers, D.~A., Bradford, C.~M., Clements, D.~L., et al.\ 2013, \nat, 496, 329. doi:10.1038/nature12050

\bibitem[Shao et al.(2019)]{shao19} Shao, Y., Wang, R., Carilli, C.~L., et al.\ 2019, \apj, 876, 99. doi:10.3847/1538-4357/ab133d
\bibitem[Shao et al.(2022)]{shao22} Shao, Y., Wang, R., Weiss, A., et al.\ 2022, \aap, 668, A121. doi:10.1051/0004-6361/202244610
\bibitem[Sharon et al.(2016)]{sharon16} Sharon, C.~E., Riechers, D.~A., Hodge, J., et al.\ 2016, \apj, 827, 18. doi:10.3847/0004-637X/827/1/18
\bibitem[Spaans(2008)]{spaans08} Spaans, M.\ 2008, EAS Publications Series, 31, 47. doi:10.1051/eas:0831010
\bibitem[Spinoglio et al.(2012)]{spinoglio12} Spinoglio, L., Pereira-Santaella, M., Busquet, G., et al.\ 2012, \apj, 758, 108
\bibitem[Uzgil et al.(2016)]{uzgil16} Uzgil, B.~D., Bradford, C.~M., Hailey-Dunsheath, S., et al.\ 2016, \apj, 832, 209. doi:10.3847/0004-637X/832/2/209
\bibitem[Valentino et al.(2021)]{valentino21} Valentino, F., Daddi, E., Puglisi, A., et al.\ 2021, \aap, 654, A165. doi:10.1051/0004-6361/202141417
\bibitem[Vallini et al.(2019)]{vallini19} Vallini, L., Tielens, A.~G.~G.~M., Pallottini, A., et al.\ 2019, \mnras, 490, 4502. doi:10.1093/mnras/stz2837

\bibitem[van der Werf et al.(2010)]{van10} van der Werf, P.~P., Isaak, K.~G., Meijerink, R., et al.\ 2010, \aap, 518, L42 
\bibitem[Venemans et al.(2016)]{venemans16} Venemans, B.~P., Walter, F., Zschaechner, L., et al.\ 2016, \apj, 816, 37. doi:10.3847/0004-637X/816/1/37
\bibitem[Venemans et al.(2018)]{venemans18} Venemans, B.~P., Decarli, R., Walter, F., et al.\ 2018, \apj, 866, 159. doi:10.3847/1538-4357/aadf35

\bibitem[Venemans et al.(2020)]{venemans20} Venemans, B.~P., Walter, F., Neeleman, M., et al.\ 2020, \apj, 904, 130. doi:10.3847/1538-4357/abc563
\bibitem[Villarreal Hern{\'a}ndez \& Andernach(2018)]{villarreal18} Villarreal Hern{\'a}ndez, A.~C. \& Andernach, H.\ 2018, arXiv:1808.07178. doi:10.48550/arXiv.1808.07178

\bibitem[Walter et al.(2003)]{walter03} Walter, F., Bertoldi, F., Carilli, C., et al.\ 2003, \nat, 424, 406 
\bibitem[Walter et al.(2022)]{walter22} Walter, F., Neeleman, M., Decarli, R., et al.\ 2022, \apj, 927, 21. doi:10.3847/1538-4357/ac49e8
\bibitem[Wang et al.(2008)]{wang08} Wang, R., Carilli, C.~L., Wagg, J., et al.\ 2008, \apj, 687, 848. doi:10.1086/591076
\bibitem[Wang et al.(2011)]{wang11} Wang, R., Wagg, J., Carilli, C.~L., et al.\ 2011, \apjl, 739, L34. doi:10.1088/2041-8205/739/1/L34
\bibitem[Wang et al.(2013)]{wang13} Wang, R., Wagg, J., Carilli, C.~L., et al.\ 2013, \apj, 773, 44. doi:10.1088/0004-637X/773/1/44
\bibitem[Wang et al.(2016)]{wang16} Wang, R., Wu, X.-B., Neri, R., et al.\ 2016, \apj, 830, 53. doi:10.3847/0004-637X/830/1/53
\bibitem[Wang et al.(2019)]{wangf19} Wang, F., Wang, R., Fan, X., et al.\ 2019, \apj, 880, 2. doi:10.3847/1538-4357/ab2717
\bibitem[Wei{\ss} et al.(2005)]{weiss05} Wei{\ss}, A., Walter, F., \& Scoville, N.~Z.\ 2005, \aap, 438, 533 
\bibitem[Wei{\ss} et al.(2007)]{weiss07} Wei{\ss}, A., Downes, D., Neri, R., et al.\ 2007, \aap, 467, 955. doi:10.1051/0004-6361:20066117
\bibitem[Willott et al.(2010)]{willott10} Willott, C.~J., Delorme, P., Reyl{\'e}, C., et al.\ 2010, \aj, 139, 906. doi:10.1088/0004-6256/139/3/906
\bibitem[Willott et al.(2015)]{willott15} Willott, C.~J., Bergeron, J., \& Omont, A.\ 2015, \apj, 801, 123. doi:10.1088/0004-637X/801/2/123
\bibitem[Wolfire et al.(2022)]{wolfire22} Wolfire, M.~G., Vallini, L., \& Chevance, M.\ 2022, arXiv:2202.05867
\bibitem[Yang et al.(2013)]{yang13} Yang, C., Gao, Y., Omont, A., et al.\ 2013, \apjl, 771, L24. doi:10.1088/2041-8205/771/2/L24

\bibitem[Yang et al.(2016)]{yang16} Yang, C., Omont, A., Beelen, A., et al.\ 2016, \aap, 595, A80. doi:10.1051/0004-6361/201628160

\bibitem[Yang et al.(2017)]{yang17} Yang, C., Omont, A., Beelen, A., et al.\ 2017, \aap, 608, A144
\bibitem[Yang et al.(2019)]{yang19} Yang, J., Venemans, B., Wang, F., et al.\ 2019, \apj, 880, 153. doi:10.3847/1538-4357/ab2a02
\bibitem[Zanella et al.(2018)]{zanella18} Zanella, A., Daddi, E., Magdis, G., et al.\ 2018, \mnras, 481, 1976. doi:10.1093/mnras/sty2394
\end{thebibliography}
\end{document}